\documentstyle[aps,preprint,tighten,graphicx]{revtex}

\begin{document}

\def\Journal#1#2#3#4{{#1} {\bf #2}, #3 (#4)}

\def\NP{{\em Nucl. Phys.} }
\def\NC{{\em Nuovo Cim.} }
\def\PL{{\em Phys. Lett.} }
\def\PR{{\em Phys. Rev.} }
\def\PRL{{\em Phys. Rev. Lett.} }
\def\PREP{{\em Phys.  Rep.} }
\def\AP{{\em Ann. of Phys.} }
\def\ZP{{\em Z. Phys.} }

\newcommand{\mathbold}[1]{\mbox{\protect\boldmath $\displaystyle #1$}}

\def\rme{{\rm e}} \def\rmi{{\rm i}}

\def\intm1p1{{\int \limits_{-1}^{1}}} \def\sumlUNE{{\sum
    \limits_{l=0}^{\infty}}}

\newcommand{\einh}{\frac{1}{2}} \newcommand{\dreih}{\frac{3}{2}}
\newcommand{\be}{\begin{equation}} \newcommand{\ee}{\end{equation}}
\newcommand{\bea}{\begin{eqnarray}} \newcommand{\eea}{\end{eqnarray}}

\newcommand{\nres}[4]{#1_{#2#3}(#4)}

\newcommand{\lmnr}[2]{{\cal L}_{\varphi N #1}^{#2}}

\newcommand{\real}{{\rm Re}} \newcommand{\imag}{{\rm Im}}

\newcommand{\Iba}{b a}

\newcommand{\bsl}[1]{#1 \!\!\! /}

\newcommand{\abs}[1]{\left | #1 \right |}

\newcommand{\dop}[1]{#1^{\prime}} \newcommand{\hatdop}[1]{\hat #1^{\,
    \prime}}

\newcommand{\solid}{\protect\rule[1mm]{6mm}{.1mm}}
\newcommand{\dash}{\protect\rule[1mm]{2mm}{.1mm}\hspace{2mm}
\protect\rule[1mm]{2mm}{.1mm}}
\newcommand{\shortdash}{\protect\rule[1mm]{1mm}{.1mm}\hspace{4mm}
\protect\rule[1mm]{1mm}{.1mm}}
\newcommand{\dashdot}{\protect\rule[1mm]{2mm}{.1mm}\hspace{2mm}$\cdot\cdot$}
\newcommand{\dashsdot}{\protect\rule[1mm]{2mm}{.1mm}\hspace{2mm}$\cdot$}
\newcommand{\dotdot}{\protect$\cdot\cdot\cdot$}

\newcommand{\FullBox}{\protect\rule[0.5mm]{2.5mm}{2.5mm}}

\pagestyle{plain} \pagenumbering{arabic}

\begin{center}
  {\LARGE A unitary model for meson-nucleon scattering \footnote {Work
      supported by BMBF, GSI Darmstadt and the U.S. DOE}\footnote
    {This paper forms part of the dissertation of T. Feuster}}
  
  \bigskip \bigskip
  
  {T. Feuster$^1$\footnote
    {e-mail:Thomas.Feuster@theo.physik.uni-giessen.de} and U.
    Mosel$^{1,2}$}
  
  \bigskip \bigskip

  {\it $^1$Institut f\"ur Theoretische Physik, Universit\"at Gie{\ss}en\\
    D--35392 Giessen, Germany\\
    \bigskip
    $^2$ Institute for Nuclear Theory, \\
    University of Washington, Box 351550, Seattle, WA98195, USA}\\[3mm]
  UGI-97-13
\end{center}

\bigskip \bigskip

%
%

\section*{Abstract}
In an effective Lagrangian model employing the $K$-matrix
approximation we extract nucleon resonance parameters. To this end we
analyze simultaneously all available data for reactions involving the
final states $\pi N$, $\pi\pi N$, $\eta N$ and $K \Lambda$ in the
energy range $m_N + m_{\pi} \le \sqrt s \le 1.9$ GeV. The background
contributions are generated consistently from the relevant Feynman
amplitudes, thus significantly reducing the number of free parameters.
The sensitivity of the parameters upon the $\pi N$-partial wave
analysis and the details of the Lagrangians and form factors used are
discussed.\\[3mm]

{\it PACS}: 14.20.Gk, 11.80Gw, 13.30.Eg, 13.75.Gx\\
{\it Keywords}: baryon resonances; N(1535); unitarity; partial wave
analysis, coupled channel

\maketitle

%
%

\section{Introduction}

A number of models have been proposed and used to extract information
about the excitation spectrum of the nucleon. The main problem faced
is the multitude of possible decay channels of the nucleon resonances.
A proper treatment of these requires both theoretical and numerical
effort. Furthermore, a large number of a priori unknown couplings is
introduced. These can only be estimated with some confidence, if all
available data are used. Ideally, all these data would be decomposed
into partial waves. Unfortunately, this has only been done so far for
some reaction channels, namely $\pi N \to \pi N$, $\pi N \to \pi\pi
N$, $\gamma N \to \pi N$. For the other possible channels we only have
total and differential cross section and polarization data available.

In most of the works only hadronic data are used to extract resonance
parameters \cite{cfhk79,ka84,ms92,sm95} since meson photoproduction
only allows the determination of the product of the hadronic and
electromagnetic couplings \cite{bmz95}. All these models employ
interaction potentials constructed to fulfill unitarity and
analyticity. The main difference between the models is the treatment
of the reaction channels. In \cite{sm95} all inelastic channels are
summed up in a 'generic' $\pi \Delta$ channel, whereas in \cite{ms92}
both $\pi N \to \pi N$ and $\pi N \to \pi\pi N$ data are fitted. In
other cases \cite{bdssnl97,sau96} the $\pi N \to \eta N$ data are used
and the $\pi\pi N$ decays of the resonances are approximated using a
dummy $\zeta$-meson. From the PDG values \cite{pdg96} it is clear that
higher lying resonances might also have other decay channels like $K
\Lambda$ and $K \Sigma$. So far couplings for these have not been
extracted in a multichannel calculation.

The $\nres S11{1535}$-resonance has long been of special interest
because of its large $\eta N$ branching ratio. This value of $\approx$
50\% is not well understood in structure models of the nucleon and
resonances \cite{cr94,gr96,bil97}\footnote{Capstick and Roberts
  \cite{cr94} are able to reproduce the $\pi N$ and $\eta N$ branching
  ratios but overestimate the partial decay widths by more than 50\%.
  Glozman and Riska \cite{gr96} explain the $\eta N$ branching ratio
  of the $\nres S11{1535}$ by the flavor-spin symmetry of the quark
  wave functions, whereas Bijker et al. \cite{bil97} suggest that the
  large $\eta N$ width of the $\nres S11{1535}$ ''is not due to a
  conventional $q^3$ state''.}. Recently also a description of the
$\nres S11{1535}$ as a quasi-bound $K \Sigma$-state has been put
forward \cite{kww97}. An accurate extraction of the $\nres
S11{1535}$-parameters would therefore constrain models of the
structure of the nucleon. Unfortunately, the values for the mass and
decay widths ($m_R$ = 1.526 - 1.553 GeV, $\Gamma_{\pi}$ = 20 - 84 MeV,
$\Gamma_{\eta}$ = 54 - 91 MeV) found in different works vary
drastically. As we will see, this is mainly due to the poor $\pi N \to
\eta N$ data.

To improve this situation, information from photoproduction
experiments might be used. Because of the rescattering these data
cannot be analyzed independently, but a combined model for the
hadronic and electromagnetic channels is needed. First attempts have
been made in the $\Delta$-region of pion-photoproduction
\cite{bdm89,skpn96}. In these unitarity was guaranteed by using the
$K$-matrix approximation. For higher energies mainly effective
Lagrangian models \cite{bmz95,fm97} have been used to extract
information on the product of the hadronic and electromagnetic
couplings. While these models have been rather successful, no attempt
has been made so far to describe the hadronic final state interaction
for all possible channels using {\it the same Lagrangians} as for the
photoproduction reactions.

As a first step in this direction we have developed a model for both
meson-nucleon and photon-nucleon reactions starting from effective
Lagrangians which is unitary and includes as many reaction channels as
is technically feasible. In this paper we present the results for the
resonance masses and widths as extracted from fits to the available
hadronic data. By using a speed plot technique described by H\"ohler
\cite{h93} we estimate the poles and residues of the resonances. In
this way we bypass a direct calculation of the $T$-matrix in the
complex energy plane, since the technical effort needed for an
analytic continuation of all Feyman amplitudes is beyond the scope of
this paper. Since our main interest is in the determination of the
hadronic couplings of the known resonances, we furthermore do not
search for additional states as done e.g. by Manley and Saleski
\cite{ms92}.

This paper is organized as follows: First the reactions included and
the available data will be listed. Then we give an overview of the
model used. This consists of a short discussion of the $K$-matrix
approximation and the Lagrangians needed. The results of the fits are
presented in comparison to the data and also the extracted masses and
partial widths will be discussed and compared to other works.

%
%

\section{Reactions channels and database}
\label{reactions}

The reaction channels in the energy regime up to $\sqrt s$ = 1.9 GeV,
to which we restrict ourselves in this paper are $\pi N \to \pi N$,
$\pi N \to \pi\pi N$, $\pi N \to \eta N$, $\pi N \to K \Lambda$ and
$\pi N \to K \Sigma$. In order to use as much information as possible
from these data, but at the same time keeping the model as simple as
possible, we adopt the following strategy:
\begin{itemize}
\item \mathbold {\pi N \to \pi N}: Here two widely used partial wave
  anlyses (PWA) are available. One is the older analysis by H\"ohler
  et al., the other is the latest version from the VPI group (SM95,
  \cite{sm95}). Recently (cited in \cite{bdssnl97}), H\"ohler (KA84,
  \cite{ka84}) has suggested to use the SM95 solution in the
  $S_{11}$-channel below the $\eta N$-threshold to account for new
  experimental data. We will present fits using both the KA84- and the
  SM95-PWA. This allows to check the dependence of the parameters on
  the analysis used.  Unfortunately, no error bars have been given for
  the KA84-solution.  Since the knowledge of the uncertainties is
  essential for all fitting procedures, errors have to be assigned to
  these data.  However, there is a certain arbitrariness involved in
  this assignment. For example, Batinic et al. choose an error that
  grows linearly with energy from some minimal value\cite{bdssnl97}.
  Here we use a different prescription, namely the error is calculated
  as:
  \be
  \Delta T_{\alpha}(W_i) \equiv \max (0.03 \: T_{\alpha}(W_i), 0.015) .
  \label{KA84err}
  \ee
  The main assumption behind this choice is that the errors are of the
  order of those of the SM95-data. Only then a comparison of the
  resulting $\chi^2$-values is meaningful. A change in the exact
  numbers in (\ref{KA84err}) does not have a sizable influence on the
  final parameters; it merely sets the scale for the $\chi^2$-values
  deduced from the fits.
\item \mathbold {\pi N \to \pi\pi N}: Manley and Saleski performed a
  decomposition of the available data with respect to various
  intermediate states like $\pi\Delta, \pi\nres P11{1440}$ and $\rho
  N$. In order to keep the model as simple as possible we do not treat
  all these states explicitly, but follow a more phenomenological
  approach \cite{bdssnl97,sau96}: the $\pi\pi N$-decay is
  parameterized by the coupling to a scalar, isovector $\zeta$ meson
  with mass $m_{\zeta} = 2 m_{\pi}$. We have chosen isovector instead
  of isoscalar (as in \cite{sau96}), to allow also decays of the $I =
  \dreih$-resonances. To determine the couplings from the results of
  Manley and Saleski we use their total $\pi N \to \pi\pi N$ cross sections
  for the different partial waves.
\item \mathbold {\pi^- p \to \eta n}: Measurements of the total and
  differential cross sections have been performed by several groups
  over a wide energy range. Unfortunately, some of these measurements
  do not agree very well with each other. Batinic et al.
  \cite{bdssnl97} have proposed a scheme to incorporate these
  discrepancies by enlarging the error for some of the datapoints.
  This scheme has also been used here. As will be seen, the large
  uncertainties in the data for this channel prohibit a good
  determination of the $\nres S11{1535}$-parameters and the $\eta
  N$-scattering length.
\item \mathbold {\pi^- p \to K^0 \Lambda, K \Sigma}: These channels
  are of minor importance over the whole energy range. Only the $K
  \Lambda$ gives a significant contribution to the total inelastic
  cross section around 1.7 GeV. Therefore we include only this
  reaction in our work. The observables used are the total and
  differential cross sections and $\Lambda$-polarizations. Due to the
  large errors the latter play only a minor role and are included for
  completeness only. A detailed description of all channels having
  strange particles in the final states is not possible anyway, since
  we have a coupling to the hyperon spectrum through $u$-channel
  contributions in this case. A determination of the parameters of the
  hyperon resonances is clearly beyond the scope of this work because
  it would require the inclusion of other reactions like $K N \to K
  N$.
\end{itemize}

Neglected are channels that lead to final states containing more than
2 pions (e.g. $\pi N \to \omega N \to \pi \pi \pi N$). In their
analysis Manley and Saleski found missing inelasticity only for some
resonances. They described this by introducing effective $\omega N$-
and $\rho \Delta$-channels that lead to 3-pion final states.
Therefore, the partial widths extracted there can only be viewed as
upper bounds for these additional decay channels. In our case only the
$\nres P13{1720}$ is affected by this. As will be discussed in Secs.
\ref{datapinpipin} and \ref{rescoupl}, we do not treat these
additional channels explicitly, but rather fit the parameters of this
resonance without the $\pi \pi N$ data.

%
%

\section{The $K$-matrix approximation}
\label{compmodels}

To solve the coupled Bethe-Salpeter equations encountered in
meson-nucleon scattering a number of models have been proposed. For
completeness we only give a short summary of the three most important
ones. The reader is referred to the references given for a more
detailed discussion.

{\it 1.} In the widely used ansatz from Cutkosky et al. (the so-called
Carnegie-Mellon Berkeley or CMB group, also used by Batinic et al.)
\cite{cfhk79,bdssnl97} the $T$-matrix in a given channel is
represented by a sum over the contributions from all intermediate
particles. The coupling $f(s)$ from the asymptotic states to these
particles determines the imaginary part of the phase factor $\Phi
(s)$:
\bea
T_{ab} &=& \sum\limits_{i,j}^N f_a(s) \sqrt \rho_a 
\gamma_{ai} G_{ij} (s) \gamma_{jb} \sqrt \rho_b f_b(s) \nonumber\\
\imag \Phi_a(s) &=& \left [ f_a (s) \right ]^2 \rho_a ,
\label{cmb}
\eea
with $\rho_a = q_a / \sqrt s$. The real part of $\Phi_a(s)$ is then
calculated from dispersion relation to ensure analyticity. With this
phase factor the self energy $\Sigma (s)$ and the dressed propagator
$G (s)$ are computed:
\bea
\Sigma_{kl} (s) &=& \sum\limits_a 
\gamma_{ka} \Phi_a (s) \gamma_{al} \nonumber\\
G_{ij} (s) &=& G_{ij}^0 (s) + \sum\limits_{k,l}^N G_{ik}^0 (s)
\Sigma_{kl} (s) G_{lj}^0 (s) . \eea
The $\gamma_{ab}$ are the free coupling parameters that are fit to the
data. Besides the known resonance contributions to $T_{ab}$ the
background is included as additional terms with poles below the $\pi
N$ threshold. The number of background parameters is therefore
proportional to the number of orthogonal channels included in the
calculation.

One of the advantages of this formalism is that it is straightforward
to search for the complex poles of the $T$-matrix since the the
potential is separable and depends only on $s$. As inelastic channels
$\eta N$, $\rho N$, $\pi \Delta$, $\pi \nres P11{1440}$, $\epsilon N$,
$\omega N$ and $\rho \Delta$ have been taken into account. Furthermore
information on the $\eta N$ threshold production amplitude was used in
the fits.

{\it 2.} In the work of Manley and Saleski \cite{ms92} the starting point is the
$S$-matrix which is written as a product of background and resonant
terms:
\bea
S &=& S_R^T S_B S_R \nonumber \\
S_B &=& \frac{1 + \rmi K_B}{1 - \rmi K_B} , \quad S_R =
\prod\limits_k^N S_k^{1/2} . \eea
Here the $S_k^{1/2}$ describes the contribution of the $k$th resonance
and is related to the $T$-matrix by:
\be S_k^{1/2} = 1 + (\rmi - x_k + (1 + x_k^2)^{1/2}) T_k, \quad S_k =
1 + 2\rmi T_k , \ee
which in turn is assumed to have a Breit-Wigner form. The n-channel
background $K_B$ is parameterized in terms of $n$ independent linear
functions of the energy $\sqrt s$. Here the inelastic channels
considered are the same as in the model of Cutkosky et al..

{\it 3.} The $K$-matrix approximation consists of choosing $K = V$ instead of
the full Bethe-Salpeter equation \cite{sau96,pj91}:
\bea
K &=& V + V \real (G_{BS}) K \nonumber \\
T &=& K - \rmi K \imag (G_{BS}) T . \eea
This corresponds to a special choice for the Bethe-Salpeter propagator
$G_{BS}$ ($k_N$ and $k_m$ are the nucleon and meson four-momentum,
respectively):
\be G_{BS} = - 2 \rmi (2 \pi)^2 m_N \; \delta (k_N^2 - m_N^2) \delta
(k_m^2 - m_m^2) \theta (k^0_N) \theta (k^0_m) ({\bsl k}_N + m_N) \ee
and leads to a rather simple equation for $T$, namely
\be T = \frac{V}{1 - \rmi V} .
\label{kmeqn}
\ee
Here no further constraints on the potential $V$ are necessary. The
simple form of (\ref{kmeqn}) makes the $K$-matrix approximation most
suitable for computation.

As stated in the introduction, we want to construct our interaction
potential $V$ starting from effective Lagrangians that describe the
couplings between different particles. The main advantage of this
ansatz is that the background contributions are calculated from the
same Feynman diagrams as the resonant amplitudes. This reduces the
number of parameters needed to describe the nonresonant background
drastically, since it is now only proportional to the number of
diagrams from which the background is determined. It is also
straightforward to incorporate various aspects like chiral symmetry by
choosing the proper $\pi N$-Lagrangian.

The main drawback is that the special choice for $G_{BS}$ used in Eqn.\
(\ref{kmeqn}) violates analyticity. Because of the more complicated
functional form of $V$ in the effective Lagrangian ansatz it is not an
easy task to restore analyticity by the use of dispersion relation
integrals (as is done in the CMB ansatz). Since the aim of this paper
is to serve as a basis for further investigations using effective
Lagrangians we do not attempt to go beyond the $K$-matrix
approximation here.

In order to test the $K$-matrix approximation Pierce and Jennings
\cite{pj91} fitted the $\pi N$-phase shifts also using other
intermediate propagators but found no significant differences in the
extracted parameters. It thus seems that all the physically relevant
contributions are already contained in (\ref{kmeqn}).

%
%

\section{Description of the model}

In an effective Lagrangian model the potential $V$ is specified in
terms of couplings between different particles. In our case these are
the nucleon, $\Lambda$, nucleon resonances and mesons. We take into
account $s$-, $u$- and $t$-channel contributions\footnote{In principle
  there is the problem of 'double counting' if one includes all
  resonances in the $s$-channel along with all $t$-channel diagrams.
  The assumption is that the relatively small number of contributions
  taken into account in the $t$-channel minimizes double counting.
  The validity of this assumption can only be investigated in a
  quantitative way, once dispersion relations are considered. This has
  to be left open for further investigations.} which can be
represented by the usual Feynman diagrams. Only in the case of $K
\Lambda$ we disregard the $u$-channel contributions since these would
come from hyperon resonances which we do not include. As mentioned
above, in this framework the background can easily be identified with
all diagrams that do not involve nucleon resonances. This limits the
number of free parameters considerably and furthermore gives
additional constraints on the resonance parameters, since the
backgrounds of the individual partial waves are no longer independent
of each other.

In this work we limit ourselves to partial waves with spin $\einh$ and
$\dreih$. We include all corresponding nucleon resonances, except for
the $\nres P31{1750}$ which has a status of only one star
\cite{pdg96}. Only for these the Lagrangians can be given in an
unambiguous way \cite{nb80,n81}, even though we already have to
include additional parameters to describe the offshell-couplings in
the case of spin-$\dreih$-resonances. Because we cannot account for
contributions of higher partial waves to total and differential cross
sections, we are limited to an energy range $\sqrt s \le$ 1.9 GeV.
This value was chosen to allow the fit of both flanks of all nucleon
resonances with spin $\einh$ and $\dreih$ to the data. Fortunately, the
resonances omitted here ($\nres D15{1675}$ and $\nres F15{1680}$) are
known to have only a small branching ratio into the $\eta N$ and $K
\Lambda$ channels \cite{bdssnl97,sz89}, so that they do not have a
strong influence on the fits to the $\pi^- p \to \eta n$ and $\pi^- p
\to K^0 \Lambda$ data.

\subsection{Background contributions}

It is well known \cite{ew88} that the $\pi N$-scattering length can be
described in the linear $\sigma$-model \cite{gl60}. There chiral
symmetry is guaranteed by inclusion of the scalar, isoscalar
$\sigma$-meson. The couplings of the $\pi$ and $\sigma$ to the nucleon
are fixed and depend only on the nucleon mass and the pion
decay-constant. In this work we use the non-linear $\sigma$-model for
guidance in constructing the coupling terms because of two reasons: i)
the $\sigma$-meson is not observed in nature, ii) in the linear model
additional terms are needed to fulfill the low-energy theorems of
pion-photoproduction \cite{bmz95,sau96} because it has pseudoscalar
(PS) instead of pseudovector (PV) $\pi N$-coupling. The coupling of
the nucleons and the pseudoscalar mesons to the vector mesons can then
be obtained by introducing the latter as massive gauge particles
\cite{kkw97}. In addition to the vector coupling we also include the
$\rho NN$ tensor coupling. As in other effective Lagrangian approaches
this mimics the breaking of chiral symmetry \cite{bmz95}. Besides
these couplings we also have the contributions from other scalar
($a_0$) and vector ($K^*$) mesons so that the total Lagrangian for the
nonresonant contributions is (suppressing isospin-factors here and in
the following):
\bea {\cal L}_{NR} = &-& \frac{g_{\varphi NN}}{2 m_N} \bar N \gamma_5
\gamma_{\mu} (\partial^{\mu} \varphi) N
- g_{sNN} s (\bar N N) 
- g_{s\varphi \varphi} s (\varphi^* \varphi) \nonumber \\
&-& g_{vNN} \bar N \left ( \gamma_{\mu} v^{\mu} - \kappa_v
  \frac{\sigma_{\mu \nu}}{4 m_N} v^{\mu \nu} \right ) N -
g_{v\varphi\varphi} \left [ \varphi \times (\partial_{\mu} \varphi)
\right ] v^{\mu}.
\label{backcoupl}
\eea
Here $\varphi$ denotes the asymptotic mesons $\pi$, $\eta$ and $K$, a
coupling to the $\zeta$-meson is not taken into account. $s$ and $v$
are the intermediate scalar and vector mesons ($a_0$, $\rho$ and
$K^*$) and $v^{\mu \nu} = \partial^{\nu} v^{\mu} - \partial^{\mu}
v^{\nu}$ is the field tensor of the vector mesons; $N$ is either a
nucleon or a $\Lambda$ spinor. For the $I = 1$-mesons ($\pi$, $\zeta$
and $\rho$) $\varphi$ and $v^{\mu}$ need to be replaced by \mathbold
{\tau \cdot \varphi} and $\mathbold {\tau \cdot v}^{\mu}$ in the
$\varphi, v NN$-couplings and by \mathbold {\varphi} and $\mathbold
v^{\mu}$ otherwise. As we will see later on, the influence of the
$a_0$ is small, whereas the $K^*$ gives the dominant contribution to
$\pi^- p \to K^0 \Lambda$ at higher energies. The parameters used for
the mesons were taken from \cite{pdg96} and are listed in Table
\ref{mesdata}.

\subsection{Resonance couplings}
\label{rescouplings}

For the coupling of the spin-$\einh$-resonances to the mesons we again
have the choice of PS or PV coupling. In principle one could start
with a linear combination of both and fit the ratio PS/PV to the data.
To keep the number of parameters small, we choose PS coupling for all
negative parity resonances and PV for positive parity. For the
negative parity case this is done in accordance with the calculation
of Sauermann et al. \cite{sau96}. For positive parity states we
choose, as for the nucleon, PV rather than PS, thus circumventing the
need for additional scalar mesons to reproduce the scattering lengths.

For the $S_{11}$- and $S_{31}$-resonances we therefore have
\be \lmnr {R_{1/2}}{PS} = - g_{\varphi NR} \bar R \: \Gamma \varphi N
+ h.c.,
\label{lnp12PS} \\
\ee
and in the case of $P_{11}$ and $P_{31}$ the couplings are given by
\be
\lmnr {R_{1/2}}{PV} = - \frac{g_{\varphi NR}}{m_R \pm m_N} \bar R
\Gamma_{\mu} (\partial^{\mu} \varphi) N + h.c. ,
\label{lnp12PV} \\
\ee
with the upper sign for positive parity. The vertex-operators $\Gamma$
and $\Gamma_{\mu}$ depend on the parity of the particles involved. For
a meson with negative intrinsic parity coupling to two baryons with
positive parity (e.g. $\pi NN$) they are given by $\Gamma = \rmi
\gamma_5$ and $\Gamma_{\mu} = \gamma_5 \gamma_{\mu}$, otherwise (e.g.
$\pi N \nres S11{1535}$) we have $\Gamma = 1$ and $\Gamma_{\mu} = \rmi
\gamma_{\mu}$.

For the spin-$\dreih$-resonances the following coupling is used:
\bea \lmnr {R_{3/2}}{} &=& \frac{g_{\varphi NR}}{m_{\pi}} \bar
R^{\alpha} \Theta_{\alpha \mu} (z_{\varphi}) \Gamma
(\partial^{\mu} \varphi) N + h.c. \nonumber \\
\Theta_{\alpha \mu} (z) &=& g_{\alpha \mu} - \frac{1}{2} (1 + 2 z)
\gamma_{\alpha} \gamma_{\mu} ,
\label{lnp32}
\eea
again with a vertex-operator $\Gamma$ that is $1$ for a particle with
negative intrinsic parity and $\gamma_5$ otherwise.

The operator $\Theta_{\alpha \mu} (z)$ allows to vary the
offshell-admixture of spin-$\einh$-fields. Some attempts have been
made to fix the parameters $z$ by examining the Rarita-Schwinger
equations and the transformation properties of the interaction
Lagrangians \cite{p69,nb80}. Unfortunately, the measured
pion-photoproduction data and $\Delta N \gamma$-transition strength
cannot be explained using these results \cite{bdm89}. Therefore, we
follow Benmerrouche et al. and others who treat the $z$'s as free
parameters and determine them by fitting the data. For a detailed
discussion of the coupling of spin-$\dreih$-particles and the problems
encountered there see \cite{n81}.

\subsection{Form factors}

In order to reproduce the measured data form factors need to be
introduced. They are meant to model the deviations from the pointlike
couplings (\ref{backcoupl}) - (\ref{lnp32}) due to the quark-structure
of the nucleon and resonances. Because it is not clear a priori which
form these additional factors should have, they introduce a source of
systematical error in all models. As we have already shown for the
case of pion-photoproduction \cite{fm97}, the extracted parameters can
depend strongly on the functional form used. To check this influence
we use three different form factors in the fits: 
\bea
F_p(q^2) &=& \frac{\Lambda^4}{\Lambda^4 + (q^2 - m^2)^2} \nonumber \\
F_e(q^2) &=& \exp (- \frac{(q^2 - m^2)^2}{\Lambda^4}) \nonumber \\
F_t(q^2) &=& \frac{\Lambda^4 + (q_t^2 - m^2/2)^2}{\Lambda^4 + (q^2 -
  (q_t^2 + m^2/2))^2} .
\label{ourforms}
\eea
$m$ denotes the mass of the propagating particle, $q$ its
four-momentum and $q_t^2$ is the value of $q^2$ at the kinematical
threshold in the $t$-channel. All parameterizations fulfill the
following criteria:
\begin{itemize}
\item they are only functions of $q^2$,
\item they have no pole on the real axis,
\item $F(m^2) = 1$ .
\end{itemize}
Furthermore, $F_p$ and $F_e$ have their maximum for $q^2 = m^2$. $F_p$
resembles a monopole-factor $\Lambda^2/(\Lambda^2 + \mathbold {q}^2)$
in the non-relativistic limit; this form was also successfully used in
other calculations \cite{pj91,sau96}. Cloudy-Bag models \cite{nbl90},
on the other hand, yield form factors $\sim \exp(- c \mathbold
{k}^2)$. $F_e$ therefore can be viewed as an extension of these to
other kinematical regimes. The main difference between both form
factors is that $F_e$ falls off more rapidly than $F_p$ far away from
the resonance position. A comparison of the extracted parameters
therefore allows one to check the influence of the offshell contributions.
In contrast to $F_p$ and $F_e$ the form factor $F_t$ enhances
contributions from low energies and does not modify the threshold
amplitudes. It was used for $t$-channel exchanges only and was
constructed to preserve the connection to the chiral symmetric ansatz
of the non-linear $\sigma$-model.

In general, one would not expect to have the same value for the cutoff
$\Lambda$ for all vertices. To take all possibilities into account we
would need to perform calculations for all combinations of couplings
and form factors, allowing $\Lambda$ to vary independently for each
vertex. Since this would introduce too many free parameters, we limit
ourselves to the following:
\begin{itemize}
\item the same functional form $F$ and cutoff $\Lambda_N$ is used in
  all vertices $\pi NN$, $\eta NN$ and $K N\Lambda$,
\item for all resonances we take the same $F$ as for the nucleon, but
  different values $\Lambda_{\einh}$ and $\Lambda_{\dreih}$ for the
  cutoffs for spin-$\einh$- and spin-$\dreih$-resonances,
\item in all $t$-channel diagrams the same $F$ and $\Lambda_t$ are
  used.
\end{itemize}
The nucleon is treated differently from the resonances to honor the
special importance of the ground-state contribution to all reactions.
The resonances themselves are split up into two categories according
to their spin, since the form of the couplings is mainly determined by
the spin of the resonances, as can be seen from (\ref{lnp12PS}) -
(\ref{lnp32}). To account for the different nature of the $t$-channel
contributions the functional form and cutoff are chosen independently
from the $s$- and $u$-channel.

\subsection{Calculation of the $T$-matrix}

Once the Lagrangians and form factors are specified, we need to compute
the $K$-matrix for all reactions and from this deduce the $T$-matrix
with the help of (\ref{kmeqn}). Here we only sketch this procedure,
all formulas needed are collected in the appendix.

As in $\pi N$-scattering \cite{ew88}, we decompose the invariant
matrix element ${\cal M}_{fi}$ in the case of mesons with the same
parity in the initial and final state as
\be {\cal M}_{fi} = \bar u (\dop p, \dop s) \left ( A + B \bsl Q
\right ) u (p,s) ,
\label{mesonpos}
\ee
with $Q$ being the average of both meson four-momenta: $Q = (q + \dop
q)/2$. Since the most general case of the scattering amplitude can be
written in terms of Pauli spinors as \cite{gw64} \be F = \chi_{\dop
  s}^{\dagger} ( \tilde A + \tilde B \, \mathbold {\sigma \cdot
  \hatdop p} \; \mathbold {\sigma \cdot \hat p} ) \chi_{s}, \ee
with the known partial-wave decomposition
\bea F &=& \frac{1}{\sqrt {\mathrm q \dop q}} \, \sumlUNE \, [ l
T_{l-} + (l+1) T_{l+} ] P_l -
\rmi \mathbold {\sigma \cdot} (\mathbold {\hatdop p \times \hat p}) 
[ T_{l+} - T_{l-} ] \dop {P_l} \nonumber \\
T_{l\pm} &=& \frac{1}{2} \intm1p1 d \cos \theta \tilde A P_l(\cos \theta) -
\tilde B P_{l\pm}(\cos \theta) ,
\label{partialpos}
\eea
we can extract the $T_{l\pm}$'s by inserting the explicit
representation of the spinors and $\gamma$-matrices \cite{bd66} into
(\ref{mesonpos}). The resulting expressions for $\tilde A$ and $\tilde
B$ in terms of $A, B$ are slightly more complicated than in $\pi
N$-scattering because we also have to take into account that the
initial and/or final hadron do not need to be a nucleon. For reactions
involving mesons with different parity the procedure is similar and
the results are listed in App. \ref{decompose}.

Once the partial-wave amplitudes $T_{l\pm}$ are given it is
straightforward to extract the various observables using standard
formulas (see App. \ref{observables} and \cite{gw64}). To include all
contributions to the cross sections we have calculated the partial
waves up to $l_{max}$ = 5. In this way the convergence of the partial
wave expansion is guaranteed.

%
%

\section{Results of the fits}
\label{fitresults}

In order to check our numerics, we reproduced the analytic results of
Hachenberger and Pirner \cite{hp78} for different contributions to the
$\pi N$-amplitude and the results of Sauermann et al. \cite{sau96}.
Especially the nonresonant background needs to be checked, because
here sign errors would remain undetected. The contributions of the
resonances are easily checked. This is because for the $s$-channel
diagrams the $K$-matrix for a given reaction $i \to f$ via a channel
with quantum numbers $\alpha$ can be written as a Breit-Wigner term
\be K_{fi}^{\alpha} = \frac{- m \sqrt {\Gamma_f^{\alpha}(s)
    \Gamma_i^{\alpha}(s)}}{s - m^2} ,
\label{kmat}
\ee
which has a pole at the resonance mass. Therefore we have a
cancellation of divergent $K$-matrix elements when computing the
$T$-matrix with the help of (\ref{kmeqn}). Any error in the
computation of the $K_{fi}^{\alpha}$'s would show up as a pole in
$T^{\alpha}$. The signs of the couplings can anyway only be determined
relative to the other contributions to the same reaction.

The $\chi^2$-fits were performed using an implementation of the
Levenberg-Marquardt algorithm. The code was derived from the IMSL
routine ZXSSQ and checked against the original version. For a number
of random parameter sets the local minimum was determined and the best
of these was taken to be the global minimum. In general the parameters
have been allowed to vary in the ranges given by the Particle Data
Group \cite{pdg96}. For the offshell-parameters the range was set to
$-2 \le 1/2 (1 + 2 z) \le 2$. To further verify the final parameter
sets these were also used as starting points for a global minimization
employing two other algorithms.

In total we extracted six parameter sets, using three different form
factors at the vertices for each of the two $\pi N$-PWA's:
\begin{itemize}
\item $F_p$ for the coupling of the nucleon, resonances and the
  $t$-channel exchanges,
\item $F_e$ for the coupling of the nucleon, resonances and the
  $t$-channel exchanges, and
\item $F_p$ for the coupling of the nucleon and resonances, $F_t$ for
  the $t$-channel exchanges.
\end{itemize}
In the following the notation is such that KA84 \cite{ka84} or SM95
\cite{sm95} denote the $\pi N$ data used in the fits. Two additional
letters indicate the form factors for $s$- and $t$-channel
contributions. Thus, for example, SM95-pt denotes a fit to the
SM95-PWA with $F_p (q^2)$ for the vertices of propagating hadrons and
$F_t (q^2)$ for the $t$-channel diagrams.

Looking at the $\chi^2$-values of the fits as given in Table
\ref{chi2comp}, it seems at first glance that the use of the KA84-PWA
leads to better overall fits. But this is mainly due to the fact that
the {\it single-energy} values of SM95 scatter around the
energy-dependent solution. That the fits for KA84 and SM95 are indeed
of equal quality can be seen from the Figures and also from the very
similar values of $\chi^2/{\rm DF}$ for channels other than $\pi N$
(Tab. \ref{chi2comp}).

The scattering lengths and effective ranges we find are in general
agreement with the values obtained by other groups. This can be seen
from Table \ref{parmcomp}, where we list both parameters $a_I$ and
$r_I$ extracted from the phase shift $S_{1I}$ close to threshold 
\cite{gw64}:
\be 
\abs {\mathbold q} \left ( \frac{1}{S_{1I}} + \rmi \right )
\approx \frac{1}{a_I} + \einh r_I \abs {\mathbold q^2} . 
\ee
Here \mathbold q denotes the meson three-momentum. The deviations from
the known $\pi N$-values are due to the fact that we fit the data over
the whole energy range and do not put special emphasis on the
threshold region. Since the Born terms and the $\rho$-contribution
dominate both the threshold amplitudes and the nonresonant background,
the high-energy behavior of these terms also influences the $\pi
N$-scattering length we find. This will be discussed in detail in Sec.
\ref{nonrescoupl}. A general trend for the $\eta N$-channel is that we
find a smaller scattering length but a larger effective range. This
indicates that our $S_{11}$-partial wave does not rise as steeply as
in the other models \cite{bdssnl97,gw97}.

For a detailed comparison of the fits we will first look at the
different reaction channels and then discuss the parameters found.

\subsection{$\pi N \to \pi N$}
\label{datapinpin}

For the fits using both the KA84- and SM95-PWA all form factors lead
to a comparably good description of the data (Figs. \ref{ppi12KA84}
and \ref{ppi32KA84}). We only show all three results for the channels
$S_{11}$, $D_{13}$ and $P_{33}$ since in the other channels the
difference is even smaller. All structures present in the data are
well reproduced. From this we conclude that nearly all major
resonances in the energy range investigated were taken into account.
The only exception seems to be the $P_{31}$. Here we clearly see in
the data the contribution of a resonance with a mass of 1.9 - 2.0 GeV.
Since a reliable determination of its parameters is not possible from
the fit to one side of the resonance only, we fit this channel only up
to 1.6 GeV. The same is true in the $S_{31}$-channel where the maximum
energy fitted was 1.8 GeV. In principle one could have higher lying
resonances in all partial waves, therefore it is clear that the fits
might not reproduce the data for energies $>$ 1.8 GeV.

As a general trend, the fits seem to be better in the $S_{I1}$- and
$P_{I1}$-channels than in $P_{I3}$ and $D_{I3}$. This might indicate a
shortcoming in the description of spin-$\dreih$-resonances. Either the
use of a common shape for the form factor for spin-$\einh$ and
spin-$\dreih$ is too restrictive or we are missing contributions from
resonances with spin $\ge \frac{5}{2}$. As can be seen from Fig.\
\ref{pics31comp}, the spin-$\dreih$-resonances give relevant
contributions to spin-$\einh$-channels away from the mass-shell. These
can be varied by changing the value of the $z$-parameters from
(\ref{lnp32}), but not totally suppressed. The same might in turn be
true for resonances with higher spin. At this point we cannot safely 
distinguish between the two explanations. 

It is interesting to note the systematics of the deviations from the
data: below the resonance it seems that we underestimate the resonance
contribution (eg. $\nres D13{1520}$, Fig.\ \ref{ppi12KA84}), whereas
for energies above the resonance position the contribution does not
fall off strongly enough (eg. $\nres P33{1232}$, Fig.\ \ref{ppi32KA84}).
This might indicate that a form factor that is asymmetric around the
resonance position might lead to a better description of the data.
Such a parameterization would then be closer to the widely used form
factors that depend on the meson three-momentum \mathbold q:
\be
F_q = \left ( \frac{\Lambda^2 + \mathbold q_R^2}
{\Lambda^2 + \mathbold q^2} \right )^{\alpha} .
\label{dipolforms}
\ee
First tests with a possible generalization of (\ref{dipolforms}) show
that this is indeed the case and that the parameters of the
spin-$\dreih$-resonances might be extracted more reliably.

In summary we find, that we can reproduce both PWA's equally well
within our model. The small differences between the two (eg. $S_{11}$
for energies $\approx$ 1.55 GeV) lead to slightly different resonance
parameters, but the systematic error induced by that is smaller than
the one coming from the different form factors used.

\subsection{$\pi N \to \pi\pi N$}
\label{datapinpipin}

Not surprisingly, the $\chi^2$-values we find for the different
reactions (Table \ref{chi2comp}) clearly show that the $\pi N \to
\pi\pi N$-channel gives the largest contribution to the total
$\chi^2$. Nevertheless, it is important to check for unusual
discrepancies in specific partial waves, because these might indicate
that resonances are missing in our calculation.

Despite the simple approximation of the two-pion state by an effective
$\zeta$-meson we find generally good fits to the partial cross
sections (Figs. \ref{p2KA84} and \ref{p2SM95}). This guarantees that
the main source of inelasticity is taken into account properly.

The exception is the $P_{13}$-channel, where we are not able to
reproduce the data at all. According to Manley and Saleski the cross
section opens up at about 1.7 GeV, but the inelasticity (as deduced
from the $\pi N \to \pi N$-data) is much larger already for energies
below that. Since this is the only resonance that exhibits this
behavior, we chose not to introduce a new reaction channel, but to fit
the $P_{13}$ parameters without the $\pi\pi N$ data. The coupling of
the $P_{13}(1720)$-resonance to $\pi\pi N$ is therefore determined by
the inelasticity in the $\pi N$-channel alone. It is thus remarkable
that the calculated $\pi N \to \zeta N$ cross section exhausts all of
the inelastic cross section, at least up to $\approx$ 1.75 GeV.

A large inelastic cross section (as deduced from the KA84-/SM95-data)
could in principle also stem from decays into other final states, but
these cannot be $\eta N$ or $K \Lambda$, because in this case we would
not be able to fit the corresponding data from $\pi^- p \to \eta n$
and $\pi^- p \to K^0 \Lambda$. Manley and Saleski indeed assumed a
coupling of a second $P_{13}$-resonance ($\nres P13{1879}$) to the
$\omega N$-channel to account for a 3$\pi$-decay. The choice of this
additional channel is, however, arbitrary, since in principle also
other decays (e.g. $\rho \Delta$) could contribute.

Unfortunately, there are already differences between the inelastic
cross sections (defined in App. \ref{observables}) as determined from
KA84 and the $\pi N \to \pi\pi N$ data as given by Manley and Saleski (e.g.
in the $S_{31}$- and $D_{33}$-channels). Especially for the $I =
\dreih$-channels this is clearly a model independent problem in the
data analyses, since there is no other reaction channel in this energy
range.

\subsection{$\pi^- p \to \eta n$}

All parameter sets give similar fits to the total and differential
cross sections (see Figs. \ref{pekKASM} and \ref{peKASM}) and the
partial waves\footnote{To avoid confusion, we plot $T^{\einh}_{\pi
    \eta}$ and $T^{\einh}_{\pi K}$ in the usual notation $\langle b |
  T_{\Iba} | a_i \rangle = \tau_i T^{\einh}_{\Iba}$ \cite{ew88}
  instead of the one given App. \ref{decompose}.} (Fig.\ 
\ref{pei12KASM}). Starting from about 1.65 GeV on upwards we find that
we cannot fully reproduce the falloff in forward direction (Fig.\ 
\ref{peKASM}). Batinic et al. \cite{bdssnl97} are able to describe the
differential data over the whole energy range, but require additional
$S_{11}$- and $P_{11}$-resonances with sizeable $\eta N$-coupling.
Unfortunately, most of the data at higher energies are from Brown et
al. \cite{bro79}, for which the uncertainties are largest. Despite
this fact the $\pi^- p \to \eta n$ reaction might be a suitable
channel to search for resonances with a weak coupling to $\pi N$. To
investigate this in detail, we would need to enlarge the energy range
of our fits to be able to extract parameters for resonances with a
mass of 1.9 - 2.0 GeV reliably. With 5 - 6 resonances coupling to this
channel, better differential data and also polarization observables
would be needed, to disentangle their contributions safely.


The agreement in the calculated partial waves between the different
fits is quite good. The discrepancies in the $P_{11}$-channel are
readily explained by small changes in the nearly vanishing coupling of
the $\nres P11{1710}$ to the $\pi N$-channel. Because of the smallness
of this coupling, the fits easily differ by 100\% for the exact value.

That the available data (esp. with the weights given by Batinic et
al.) do not put too strong constraints on the couplings can be seen
best when looking at the total cross sections (Fig.\ \ref{peKASM}).
Even though these show sizable deviations from each other above 1.65
GeV, all lead to a rather similar $\chi^2$-values in this channel.

\subsection{$\pi^- p \to K^0 \Lambda$}

As in the case of $\pi^- p \to \eta n$ inconsistencies between
different measurements of the cross sections can be observed (e.g. at
1.694 GeV in Fig.\ \ref{pkKASM}). Also the errors of the polarization
data given in \cite{sax80} are extremely large. In practice these data
do not constrain the couplings at all. So also in this channel better
data are needed. The contribution to the total $\chi^2$ is larger for
this channel than for the $\eta$-production (Table \ref{chi2comp}).
This is mainly due to the fact that we did not enlarge the errors as
in the case of $\pi^- p \to \eta n$.

In Fig.\ \ref{pki12KASM} we also show the partial waves extracted from
our calculations together with the results of Sotona and \v{Z}ofka
\cite{sz89}, obtained in an analysis of $\pi^- p \to K^0 \Lambda$
alone. Since we find an appreciable coupling to the $K
\Lambda$-channel only for two resonances ($\nres S11{1650}$ and $\nres
P11{1710}$), all our fits yield very similar partial waves. In
contrast to this, the values from Sotona and \v{Z}ofka differ strongly
from our results. Nevertheless, for the lower energies up to about
1.8 GeV both models describe the experimental data equally well.
This shows the importance of coupled channel analyses, since the data
for the $\pi^- p \to K^0 \Lambda$ reaction alone obviously do not
allow to determine the partial waves (and thus the resonance
parameters) uniquely.

We stress again that we do not include all contributions to $\pi^- p
\to K \Lambda$ in our analysis. As already pointed out in Sec.
\ref{reactions}, hyperon-resonances are omitted and therefore
$u$-channel contributions are missing in the calculation. Furthermore,
the rescattering through a $K \Sigma$ intermediate state might change
the angular distribution. The influence of this additional channel can
be seen in Fig.\ \ref{pekcomp}, where we also show the results of
Kaiser et al. \cite{kww97} for the total $\pi^- p \to K^0 \Lambda$
cross section. In their calculation the cusp due to the opening of the
$K \Sigma$-channel at 1.68 GeV is clearly visible.

Keeping this in mind we find, that the fits account for most of the
data. Only for the highest energies considered there are indications
for additional contributions from resonances omitted here (see Fig.\
\ref{pickscomp}, right). For the good overall quality of the fit the
$K^*$-meson is essential, as can be seen from Fig.\ \ref{pickscomp}.
For the higher energies the forward peaking is solely due to this
$t$-channel contribution. At the same time the influence on the other
angles is small so that the resonance couplings can still be
determined quite accurately.

%
%

\section{Parameters and couplings}

From the detailed discussion in the last section it is evident that a
simultaneous description of all available data is possible within this
model. The main resonances and the dynamical rescattering seem to be
incorporated correctly; therefore reliable parameter estimates are
possible. The results of these are given in this section. We thus now
turn to the discussion of the couplings found in the various fits,
starting with the background parameters. As already pointed out, the
nonresonant background is made up by a few Feynman diagrams only and
can therefore not be varied independently for each channel. As a
consequence the extraction of the resonance parameters depends
strongly on the quality of the 'overall fit'. This will be made clear
in more detail at the end of this section.

In general we find that the systematical error that can be deduced
from fits with different form factors and/or data sets is more
important than the statistical error found in each fit. We therefore
do not give any statistical errors in the various tables.

\subsection{Meson nucleon couplings}
\label{nonrescoupl}

The couplings of the mesons to the nucleon, as determined in the fits,
are listed in Tables \ref{mesparmKA84SM95} and \ref{mescoupl}. To
exhibit the influence of the form factor of the nucleon and the
$t$-channel exchanges, we both show the couplings at the onshell-point
$\sqrt s = m_N$ and at the thresholds of the $s$- and $t$-channel,
respectively (Table \ref{mescoupl}). Furthermore, we list the
cutoff-parameters $\lambda_{N,\einh,\dreih,t}$ in Table
\ref{cutoffsKA84SM95}.

For the couplings to $\pi$, $\eta$ and $K$ (a $\zeta NN$-vertex was
not taken into account) we in general find that our values are
somewhat lower than those obtained by other groups. Furthermore, we
observe only a small spreading of the values for $g_{\pi NN}$ from the
different fits, which indicates the important role of the Born terms
for the $\pi N$ nonresonant background. For the other couplings
($g_{\eta NN}$ and $g_{K N \Lambda}$) this is not the case, mainly
because the form factors $F_{p,e}$ lead to a large reduction of these
contributions ($F_{p,e} \approx 0.3 - 0.7$ at threshold). Even with
the couplings set to zero, we would still be able to reproduce the
$\pi N \to \eta N$ and $\pi N \to K \Lambda$ data with only a minor
increase of $\chi^2$. This indicates that these processes are
determined by $t$-channel and resonance excitations. In
meson-photoproduction the situation is different, because the
requirement of gauge invariance counteracts the influence of the form
factor \cite{sau96,bmk97}. Therefore, in these reactions one might be
able to extract the $g_{\eta NN}$ and $g_{K N \Lambda}$ couplings more
reliably.

Since the nonresonant background in this model is made up from the
Born terms and the $t$-channel exchanges, it is completely determined
by a relatively small number of parameters. In particular it cannot be
varied independently in different partial waves, as for example in
\cite{ms92,bdssnl97,dvl97}. Therefore, constraints on the background
found in one channel might influence all other extracted parameters.
This provides a stringent test of the model that is not possible in
other works.

To illustrate the coupling between background and resonance parameters
we look at the $t$-channel contribution of the $\rho$-mesons to $\pi
N$-scattering. The $t$-channel $\rho$-exchange leads to the following
amplitudes \cite{hp75}:
\bea {\cal M}_{fi} &=& \bar u (\dop p, \dop s) \left ( A + B \bsl Q
\right )
u (p,s) \nonumber\\
A &=& \frac{g_{\rho NN} \kappa_{\rho NN} g_{\rho \pi\pi}}{2 m_N} 
\frac{s - u}{t - m_{\rho}^2} \cdot F(t) \nonumber \\
B &=& 2 g_{\rho NN} (1 + \kappa_{\rho NN}) g_{\rho \pi\pi} \frac{1}{t
  - m_{\rho}^2} \cdot F(t) .
\label{rhocontr}
\eea
Since $(s - u)/(t - m_{\rho}^2)$ diverges with energy, this
contribution will dominate all others from some point. In order to
reduce the divergent increase of $A$ from (\ref{rhocontr}) with energy
the fits drive the effective couplings $g \cdot F$ down by reducing
$g$.

This effect can be seen best for $\rho$ and $K^*$. With small
couplings $g_{\rho NN}$ and $g_{K^* N\Lambda}$ the fit is improved for
the highest energies considered, but at the same time leads to a too
small background for the lower energies. As a consequence we find
systematic deviations for example in the $P_{33}$-channel at around
1.4 GeV (comp. Figs. \ref{ppi32KA84} and \ref{ppi32SM95}). This in
turn causes the small values for mass and width of the
$\Delta$-resonance. From this it is clear that we need a stronger
modification of the $\rho$- and $K^*$-contribution even for energies
below 2.0 GeV to have the desired Regge-like behavior (e.g. as in
\cite{cfhk79}). This could possibly be achieved by a form factor that
is a function of all three variables $s$, $u$ and $t$ and can,
therefore, at best be approximated by our choices for $F_p$, $F_e$ and
$F_t$. For the $a_0$ the situation is not so clear, since it is a
scalar meson and does not give a divergent contribution to the
scattering amplitude.

The values for the tensor couplings of the $\rho$ (Table
\ref{mescoupl}) are smaller than the VMD-value of 3.71 used by
H\"ohler and Pietarinen \cite{hp75}, whereas Pearce and Jennings
\cite{pj91} deduced a value of 2.25 in a model similar to ours. It
should be noted that in \cite{hp75} two different form factors have
been used for the vector and tensor coupling of the $\rho$. Due to
this additional $t$-dependence it is not straightforward to compare
the value given there to our numbers. Furthermore, one has to keep in
mind that H\"ohler and Pietarinen used an analytic continuation of the
$\pi N$ amplitudes together with the P-wave $\pi \pi N \bar N$ phase
shifts in order to extract the $\rho NN$ vector and tensor couplings.
Therefore, one would expect to find similar values only if dispersion
relation constraints would be incorporated in our ansatz. This is
clearly one of the main points to improve in further calculations.

For the $K^*$ the tensor couplings are essentially equal in all fits 
because of the extreme sensitivity of the differential 
$\pi^- p \to K^0 \Lambda$ cross section in forward direction. This is 
shown in Fig.\ \ref{pickscomp}, where for two energies the $K^*$-meson 
contribution is turned off. In contrast to this, the coupling of the 
$a_0$ is not very well determined. This is so because there are
several nucleon-resonances with non-vanishing $\eta N$-decays (see 
Tables \ref{rescoupl1212comp} - \ref{rescoupl32comp},
\ref{rescouplKA84} and \ref{rescouplSM95}) and therefore, because of 
the stronger interference of $s$-channel amplitudes, no region exists 
where the $t$-channel contribution is dominant.\\

In general all fits yield similar couplings, especially if one
focusses on the effective values $g \cdot F$ (see Table
\ref{mescoupl}). This indicates that the nonresonant background is,
apart from the discussed vector-meson contributions at higher
energies, properly taken into account. From this we expect that the
resonance couplings also do not show large deviations between the
different fits, since the background is of comparable size.

Unfortunately, we cannot compare our nonresonant contributions to the
scattering amplitude with the results of other calculations, since the
explicit parameters used in the calculation of the background are
mostly not given \cite{ms92,bdssnl97}. Only Dytman et al. \cite{dvl97}
show the background for the case of the $S_{11}$-channel. A comparison
with our fit KA84-pt is plotted in Fig.\ \ref{pics11back}. One finds
drastic differences, even though the full amplitude is in good
agreement. Especially near threshold our amplitude is dominated by the
background, as is to be expected from chiral symmetry \cite{ew88}.
Additionally, in fit KA84-pt one notices the opening of the $\eta
N$-threshold even for the nonresonant contribution. This is due to the
$\nres D13{1520}$-resonance and its decay into $\eta N$. Both features
are not present in the calculation of Dytman et al.. This shows that a
comparison of resonance parameters obtained by groups that use an
explicit background parametrization is only meaningful, if the
background parameters are given.

\subsection{Resonance parameters}
\label{rescoupl}

In this section we discuss the masses and widths of the nucleon
resonances we have extracted. First the $I = \einh$-resonances in the
channels $S_{11}$, $P_{11}$, $P_{13}$ and $D_{13}$ and secondly the $I
= \dreih$ ($S_{31}$, $P_{31}$, $P_{33}$ and $D_{33}$) excitations will
be investigated.

For comparison we first quote the results of other analyses in Tables
\ref{rescoupl1212comp} - \ref{rescoupl32comp}. Batinic et al.
\cite{bdssnl97} only took $I = \einh$-channels into account and did
not include a coupling to $K \Lambda$. In Cutkosky et al.
\protect\cite{cfhk79}, H\"ohler et al. \protect\cite{ka84} and Arndt
et al. \cite{sm95} the $\pi N$-scattering data were used and only the
total widths and $\pi N$-branching ratios were given. Manley and
Saleski \cite{ms92} used the data from $\pi N \to \pi N$ and $\pi N
\to \pi\pi N$ in their fits; the other couplings were determined from
the missing inelasticity alone. Therefore, the numbers given for decay
channels other than $\pi N$ and $\pi \pi N$ only indicate that
additional decay channels need to be present to account for the total
inelasticity\footnote{The only exception is the $\nres S11{1535}$.
  In this case no other channel except $\eta N$ is open at the resonance
  energy.}. The different results from the various models illustrate
that only the simultaneous fit to all open reaction channels allows
the extraction of parameters for resonances with small $\pi
N$-branching fraction (e.g. the $\nres P33{1600}$, which was not found
in \cite{sm95}).

Listed in Tables \ref{rescouplKA84}, \ref{rescouplSM95} and
\ref{reszparmKA84SM95} are all masses, decay widths and $z$-parameters
for the 6 fits done. We do not list the corresponding couplings, since
a meaningful comparison to other calculations can only be done in
terms of the decay widths. The reader is referred to App. \ref{ccanddw}
for a complete list of formulas needed to extract the coupling
constants. The decay widths and branching ratios were calculated on
resonance ($\sqrt s = m_R$); since we include $q$-dependent form
factors at the corresponding vertices, the total decay widths do {\it
  not} represent the FWHM that is seen e.g. in the resonance
contribution to the total scattering cross section. In brackets we
indicate the signs of the coupling constants. These where taken to be
the same as in Manley and Saleski \cite{ms92} for the $\pi N$ and
$\pi\pi N$ decays.

\subsubsection{Isospin-$\einh$-resonances}
\label{rescoupl12}

\mathbold {S_{11}}: For this channel there are a number of detailed
models \cite{sau96,tbk94} that aim to extract the parameters of the
$\nres S11{1535}$. This resonance is of special interest because of
its large $\eta N$-branching. The deeper reason for this is not well
understood and rather different explanations have been given
\cite{cr94,gr96,bil97,kww97} (see the corresponding footnote in the
introduction). A reliable value for this parameter would therefore put
strong restrictions on all models for this resonance. Since we have at
least two resonances in this channel close to each other, a
satisfactory fit is only possible if both are included \cite{sau96}.
Furthermore the s-waves $S_{11}$ and $S_{31}$ at threshold are
dominated by the Born terms and the $\rho$-meson that determine the
scattering lengths. In addition, at least the two channels $\pi N \to
\pi N$ and $\pi N \to \eta N$ have to be taken into account because of
the large branching of the $\nres S11{1535}$ ($\approx$ 50\% $\pi N$,
$\approx$ 45\% $\eta N$) into both of these. This has two
consequences: i) only within a model accounting for all these points a
reliable determination of the $\nres S11{1535}$-parameters is possible
and ii) all extractions are limited by the quality of the $\pi N \to
\eta N$ data.

In Table \ref{rescoupl1212comp} in addition to the other values the
$S_{11}$-parameters extracted from \cite{sau96,dvl97} are given. In
the work of Sauermann et al. also the $K$-matrix approach was used,
but within the linear $\sigma$-model instead of the pseudovector $\pi
NN$-coupling and without the $\rho$-meson used here. In spite of this
the agreement in the parameters is quite good, only for the $\eta
N$-width we find some differences (95-113 MeV using KA84 as compared
to 89 MeV in \cite{sau96}) that might be related to the different form
factors used. The same holds for the other models as well. As was
already discussed in the last section, this discrepancy may be also
due to the treatment of the nonresonant background in the different
calculations.

Unfortunately, the spreading of the parameters is larger for the fits
to the SM95-PWA. This is because we were not able to reproduce 
the data for the real part of the $S_{11}$-partial wave near the
minimum at 1.55 GeV and for the maximum of the imaginary part just
above 1.5 GeV (Fig.\ \ref{ppi12SM95}). This is interestingly also the
region of the largest differences between both the KA84-PWA and the
energy-dependent solution of SM95 to the energy-independent data.
Maybe the assignment of larger error bars for these energies would lead
to more consistent values for the $\nres S11{1535}$ parameters.

For the second resonance, $\nres S11{1650}$, a comparable $\pi
N$-branching is found in all models, whereas the $\pi\pi N$-width
comes out larger in our fits. Since the $\pi\pi$-states is
approximated by a $\zeta$-meson \cite{sau96}, this does not
necessarily lead to other scattering amplitudes. Furthermore we notice
that we find no significant coupling to the $\eta N$-channel, but a
5-8 \% decay into $K \Lambda$. Such a coupling is known from
kaon-photoproduction \cite{sz89,bmk97}.

Since other models find additional $S_{11}$-resonances at 1.8 - 1.9
GeV \cite{ms92,bdssnl97}, these states might influence the couplings
of the $\nres S11{1650}$. Unfortunately, the given values for the
$\nres S11{2090}$ are not in good agreement with each other.
Therefore, no definite conclusions can be drawn about a possible
change of parameters due to this resonance.\\

\mathbold {P_{11}}: Due to the large, varying background from the Born
terms and the $\Delta$-resonance and because of its large decay width,
the mass of the $\nres P11{1440}$ cannot be determined well. Only the
branching ratios are in good agreement with the other models
(60-70\% $\pi N$, 30-40\% $\pi\pi N$). Again we find that the
parameter sets with higher mass yield larger widths. A coupling to the
$\eta N$-channel is found in all fits, but the quality of the data
does not allow a precise determination of the $\eta N$ decay width.
Since we also have the coupling of the nucleon to the $\eta$ it is
questionable if these two contributions can be fully disentangled.

In the energy range of the $\nres P11{1710}$ the $t$-channel
$\rho$-meson contribution dominates the amplitude. Therefore the
parameters of this resonance are sensitive to the form factors and 
cutoffs used and vary accordingly. Interestingly, all fits find a 
very small ($<$ 1 MeV) $\pi N$-coupling so that the contribution to the
$P_{11}$-partial wave comes solely from rescattering. This makes the 
parameters of the $\nres P11{1710}$ sensitive to the
unitarization-procedure used in the different models. The structure in
the SM95-PWA seems to indicate a much broader resonance in this energy
region. Clearly we cannot fit these data very well.\\

\mathbold {P_{13}}: All models agree that the width of the $\nres
P13{1720}$-resonance is dominated by the $\pi\pi N$-decay. The higher
mass we find in our fits is determined by the imaginary part of the
$\pi N$-phase shift. Since Manley and Saleski\cite{ms92} list another
$P_{13}$-resonance at 1.879 GeV, it is not clear if our $\nres
P13{1720}$ is some kind of average of both resonances in this
energy range. To answer this question, the fits would have to be
extended to higher energies to cover the full range of all possible
resonances.

The discrepancies to the $\pi N \to \pi\pi N$ data have been discussed
already in Section \ref{datapinpipin} and might be due to a missing
decay channel ($\omega N$, $\rho \Delta$). The spread of the
parameters is also present in the $z$-values, that differ between all
fits (see Table \ref{reszparmKA84SM95}).\\

\mathbold {D_{13}}: As already mentioned in Sect. \ref{datapinpin}, we
find systematic deviations from the $\pi N$ data for all
spin-$\dreih$-resonances. Besides for the $\Delta$ this effect is most
prominent for the $\nres D13{1520}$. The underestimation of the data
for energies around 1.4 GeV leads to a small mass in all fits. Related
to this we also find smaller values for the partial decay widths,
whereas the branching ratios are similar to the values given in Table
\ref{rescoupl1232comp}. Especially the $\eta N$-decay is noticeable.
The small width does not imply a small coupling, since the $\nres
D13{1520}$ is close to the $\eta N$-threshold at 1.49 GeV. That this
coupling can be extracted at all is due to the fact that the s-wave -
d-wave interference is responsible for the observed lack of isotropy
in the differential $\pi^- p \to \eta N$ cross section around the
$\nres S11{1535}$-resonance.

For the $\nres D13{1700}$ the results obtained by different groups
vary strongly. Whereas Manley and Saleski \cite{ms92} give parameters for
this state, it is not present any more in the latest analysis of Arndt
et al. \cite{sm95}. The same is true for our fits, where the second
$D_{13}$-resonance is found at 1.9 GeV. Since Batinic et al.
\cite{bdssnl97} find two resonances in this energy range, (at 1.817
and 2.048 GeV), the parameters given here have to be treated with the
same caution as in the case of the $\nres P13{1720}$. Furthermore, we
cannot reliably determine the parameters of the second $D_{13}$-resonance
since we only include data up to 1.9 GeV. Accordingly, we find no
agreement between the different fits for the couplings and especially
the $z$-parameters.

\subsubsection{Isospin-$\dreih$-resonances}
\label{rescoupl32}

\mathbold {S_{31}}: Our values are similar to those given by
\cite{dvl97,sm95}, whereas Manley and Saleski find the $\nres
S31{1620}$-resonance at 1.672 GeV with a $\pi N$-partial width of 9\%.
The reason for this might be found in the $\pi\pi N$-approximation
used in this work. Since Manley and Saleski find two strong channels for
the $\pi\pi N$-decay ($\pi \Delta$ $\approx$ 62\% and 
$\rho N$ $\approx$ 25\%), one cannot expect to obtain a good
description of this decay by an effective $\zeta$-meson. This problem 
is independent of the form factors used, as can be seen from the
similar values in all fits.\\

\mathbold {P_{31}}: As discussed in Sec. \ref{datapinpipin}, we do not 
include a resonance in this channel. The data are only fitted up to
1.7 GeV; within this range no resonance appears (apart from a one-star
candidate $\nres P31{1744}$ given by Manley and Saleski \cite{ms92}).

Because of this we here have an indication of how well the
non-resonant background is described in our model. For all fits we
find that we overestimate the size and the shape of the real part of
$P_{31}$ for energies $\approx$ 1.35 GeV. Since the background is
dominated by the Born terms and the $\rho$-exchange in this region, an
improvement of the description in this channel could only be achieved
by reducing the quality of the fit in some other channel(s). 

Pearce and Jennings found that the same deviations only occur within
the $K$-matrix approach and not when using other frameworks \cite{pj91}.
From this we conclude that for a better description of the data in
this channel one would need to go beyond the $K$-matrix
approximation used in this work.\\

\mathbold {P_{33}}: As expected, all fits lead to the same parameters
for the $\nres P33{1232}$. The numbers are slightly lower than in the
other works. This has already been explained in Section
\ref{nonrescoupl} by the $\rho NN$-form factor used in our
calculation, that forces a smaller $\rho NN$-coupling than usual. The
fits try to compensate for this by lowering the mass and the width of
the $\nres P33{1232}$.

The second resonance, $\nres P33{1600}$, can be clearly seen in the
$\pi N \to \pi\pi N$-channel, whereas the contribution to the $\pi
N$-phase shift is negligible. Despite the discrepancy between the
inelasticities from KA84/SM95 and the $\pi N \to \pi\pi N$-cross
section, the couplings of the $\nres P33{1600}$ are well determined
and are comparable to the values of Manley and Saleski ($m_R$ = 1.706 GeV,
$\Gamma_{tot}$ = 430 MeV).

In contrast to the $I = \einh$-case, the $z$-parameters are very well
determined for the $\nres P33{1232}$. As Fig.\ \ref{pics31comp} shows,
this is due to the strong offshell-contribution to the
$S_{31}$-partial wave. Since the offshell-part of the coupling is
governed by the $z$-parameters, the high sensitivity of the fits is
easily understood. Only a few extractions of $z_{\pi}$ of the $\nres
P33{1232}$ have been performed so far. Olsson and Osypowski
\cite{oo78} have used both $\pi N$-scattering data and
pion-photoproduction. They found $z_{\pi}$ = -0.45 ($\pi N$) and
$z_{\pi}$ = -0.29 (photoproduction).  In another analysis of $\gamma N
\to \pi N$ Davidson et al. \cite{dmw91} deduced $z_{\pi}$ = -0.24. All
these values are in excellent agreement with the results of our fits
(-(0.33 - 0.38) for KA84 and -(0.31 - 0.35) for SM95), especially
since the corresponding
offshell-contributions are influenced by the rescattering.\\

\mathbold {D_{33}}: Similar to the $S_{31}$-channel we find a
resonance with weak coupling to $\pi N$. Therefore, the parameters of
the $\nres D33{1700}$ are determined by the $\pi N \to \pi\pi N$ data.
Accordingly (as for the $\nres S31{1620}$), the masses we find are
lower than the value of Manley and Saleski. As in the other cases, the
partial widths are also smaller, but the branching ratios are in good
agreement.

Again, the $z$-parameters are in good agreement between the different
fits with the exceptions of KA84-pt and SM95-ee, where we find the same
magnitude but opposite sign of $z_{\pi}$. This parameter is fixed
mainly by the large contribution of the $\nres D33{1700}$ to the
$P_{31}$ -partial wave. Since we do not include a resonance in this
channel, the value of $z_{\pi}$ depends on the interference with all
other background contributions and is therefore only well determined
with respect to all these other couplings.

\subsection{Pole positions and residues}
\label{respoles}

As we have already stated in the introduction, we do not attempt to
continue the $T$-matrix into the complex energy plane to locate the
poles. The reason is mainly a technical difficulty in the effective
Lagrangian approach. In this framework all Feynman diagrams would have
to be calculated for complex energies and then decomposed into the
partial waves. For the other models described in Sec. \ref{compmodels}
the poles can be found more easily, since there the potential $V$ is
determined in each partial wave independently and can, therefore, be
chosen to be a function of $s$ only.

As a first approximation we estimate the location of the poles of the
$T$-matrix following a method used by H\"ohler \cite{h93}. There the
so-called {\it speed} of the amplitudes is used to determine the poles
and residues directly from the PWA data. For details of the method see
\cite{h93}.

Starting point is the quantum mechanical consideration that the
formation of an unstable excited state in a reaction leads to a
time-delay $Q$ between the outgoing wave packet and an undisturbed
wave that can be calculated from the scattering amplitude
\cite{gw64,h93}:
\be 
Q = - \rmi \frac{dS}{dW} S^{-1} = 2 \abs {\frac{dT}{dW}}, 
\qquad W = \sqrt s. 
\ee 
The second equality holds for the case of elastic
scattering. This can easily be generalized to the multichannel case.
The speed is now defined as:
\be 
Sp (W) = \abs {\frac{dT}{dW}}. 
\ee
A peak of this speed in general corresponds to the formation of a
resonance state. For the $\pi N$ scattering this is the case except
for the cusp in the $S_{11}$-partial wave that is due to the opening
of the $\eta N$ decay channel. Resonance parameters can therefore
(with the exception of the $\nres S11{1535}$) also be obtained from
{\it speed plots} that show $Sp(W)$ vs. $W$.

Following \cite{h93} we now assume the $T$-matrix to be of the form
\be 
T(W) = T_{back} (W) + 
\frac{R \Gamma \rme^{\rmi \Phi}}{m_R - W - \rmi \Gamma/2}
\label{efftmat}
\ee
in the vicinity of a resonance (= maximum of $Sp(W)$). Here $m_R -
\rmi \Gamma/2$ is the location of the pole in the complex energy plane
and $R \Gamma \rme^{\rmi \Phi}$ is the residue. $T_{back} (W)$ is the
background amplitude due to nonresonant contributions. If the energy
dependence of $T_{back}$ can be neglected the speed only depends on
the resonance parameters $m_R, \Gamma, R$ and $\Phi$. Using $T_{back}
= const.$ we find:
\bea
\frac{dT}{dW} &=& \frac{R \Gamma \rme^{\rmi \Phi}}
{(m_R - W - \rmi \Gamma/2)^2} \nonumber\\
Sp(W) &=& \frac{R \Gamma}{(m_R - W)^2 + \Gamma^2/4}
\label{effspeed}
\eea
Our procedure is now as follows: first, determine $m_R, \Gamma$ and
$R$ by fitting the speed given in (\ref{effspeed}) to the calculated
partial waves and secondly, use this input to fix $\Phi$ from $dT/dW$.
In this way we can extract resonance parameters directly from the
unitarized $T$-matrix, consistent with the method usually used to
determine resonance parameters from actual data.

Since in an effective Lagrangian model all background contributions
are well determined, one might try to discard all $u$- and $t$-channel
contributions to reduce $T_{back} (W)$ in (\ref{efftmat}). This would
allow a better extraction of the resonance parameters in cases where
the background is not energy-independent. Unfortunately, due to
rescattering, this does not work in the $K$-matrix approach. Even if
we had a constant background $K_{back} (W)$ we could not disentangle
its contributions to the $T$-matrix from the resonant part.

The results of these fits are given in Tables \ref{resparmpoles12} -
\ref{resparmpoles32}, together with the values obtained in other
models. The agreement for the pole positions between the different
models is in general better than for the mass and width values listed
in Tables \ref{rescoupl1212comp} - \ref{rescoupl32comp}.

Furthermore, we note again that the decay widths extracted in our fits
and given in the Tables \ref{rescouplKA84} - \ref{rescouplSM95} are
the values at the resonance positions and that the energy-dependent
width also includes the respective form factors. In contrast to this
the imaginary part of the pole position is (in our case) the width of
a Lorentz function (\ref{effspeed}) fitted to the speeds and therefore
corresponds to the FHWM of the resonance. From this it is easy to
understand that the width deduced from the pole positions is in
general smaller than the value of the energy-dependent width on the
resonance, since our form factors decrease the resonance contributions
for energies away from the resonance mass.

For the $\nres S11{1535}$ the pole position cannot be determined from
the speed plot, since a peak due to the opening of the $\eta N$ channel
dominates in this energy region. For the $\nres D13{1700}$ and $\nres
P33{1600}$ no parameters could be extracted because they only appear
as a shoulder in the speed plots. Here maybe a fit to a speed plot
derived from the $\pi\pi N \to \pi\pi N$ elastic amplitude could be
used, since the $\pi\pi N$-decay is their major decay branch ($\approx$
85 \%). Furthermore, we find from the resulting Argand plots for
$dT/dW$ that the assumption of a constant background is not justified
in the cases of $\nres P11{1710}$, $\nres P13{1720}$, $\nres S31{1620}$
and $\nres D33{1700}$. For these resonances an analytic continuation
of the whole $T$-matrix would be needed to determine the pole
positions more reliably.

The good agreement of the parameters obtained from our model with the
results of the other models again shows the ability of the effective
Lagrangian approach to describe the data.

\subsection{Interdependences of parameters}

At the end of this discussion we focus on the interdependences of
different parameters as determined from the covariance-matrix $[C]$ of the
fits. To this end we extracted the coefficients of correlation given by:
\be
r_{ij} = \frac{C_{ij}}{\sqrt {C_{ii}C_{jj}}} .
\ee
In contrast to the covariances $C_{ij}$, the $r_{ij}$ are restricted
to values between -1 and 1 and therefore give a measure of the
correlation that is independent of the individual variances $C_{ii}$ of
the parameters. The most pronounced correlations we find for the
following cases:
\begin{itemize}
\item As to be expected, the different parameters of a specific
  resonance (like mass and width) are strongly ($\abs r \approx$ 0.6 -
  0.9) correlated with themselves. The same is true for the cases
  where we have two resonance in a partial wave.  Here we find a
  strong interdependence between the parameters of both resonances
  (esp. in the $S_{11}$- and $P_{11}$-channel, $\abs r \approx$ 0.8).
\item Also easily understood are the correlations between the
  parameters of the $S_{I1}$- and $P_{I1}$-resonances and the
  $z$-parameters of the $P_{I3}$- and $D_{I3}$-resonances. This has
  already been pointed out in Sect. \ref{rescoupl32} for the case of
  $z_{\pi}$ of the $\nres P33{1232}$ (comp. Fig.\ \ref{pics31comp}).
  The same effect can be seen for the other channels as well, even
  though the values for the $z$-parameters vary between the different
  fits.  Therefore, this effect can best be seen in the correlations
  and not in the parameters themselves. Noticeable here are the
  correlations of the $\nres S31{1620}$-parameters to the offshell
  contributions of the $\nres P33{1600}$ and the $\nres D33{1700}$.
  For the $I = \einh$-resonances the $\nres P11{1440}$-parameters
  exhibit large dependencies to the $z$-parameters of the $\nres
  P13{1720}$.
\item For the $\nres P11{1440}$ we also find a strong correlation to
  the parameters of the $\nres S31{1620}$ ($\abs r \approx$ 0.7). This
  surprising result has its explanation in the $u$-channel
  contributions of the latter to the partial wave $P_{11}$. Because
  the $\nres P11{1440}$ is a rather broad resonance its parameters are
  influenced by this background that is most important for energies
  $\approx$ 1.5 GeV.
\item Since the background is in our model given by a few
  contributions only, it is not independently fixed in the different
  partial waves. Accordingly, we find we find some degree of
  interdependence between the nonresonant parameters, mainly between
  $g_{\pi, \eta NN}$, $g_{K N \Lambda}$ and the various $z$-parameters
  of the spin-$\dreih$-resonances.
\item The parameters of the $\nres D13{1700}$ show a rather large
  correlation to the couplings of the other resonances. This indicates
  that the couplings of the $\nres D13{1700}$ are not well determined
  by the $D_{13}$-partial wave data; instead they are governed by
  offshell contributions of this resonance to the other partial waves.
  Since we find this state at the highest energies we consider in this
  work (1.9 GeV), its parameters cannot be extracted reliably.
\end{itemize}
These considerations are a further indication that the resonance
parameters (with the exception of the $\nres D13{1700}$) are
determined reliably in this model. The unexpected correlations of the
$\nres P11{1440}$ to the $\nres S31{1620}$ point to some 'hidden' form
factor dependence that is not obvious from the extracted parameters
alone.

%
%

\section{Comparison with the $T$-matrix approximation}

So far, in most models for $\gamma, \pi N \to \eta N, K \Lambda$ the
$T$-matrix approximation has been used \cite{bmz95,sz89,tbk94,bmk97}.
In this ansatz the $T$-matrix is calculated {\it directly} from the
lowest order Feynman diagrams. For the resonance contributions the
imaginary part of the amplitude is introduced by hand through the
inclusion of a width in the propagators:
\be 
T_{fi}^{\alpha} = \frac{- m \sqrt {\Gamma_f^{\alpha}(s)
    \Gamma_i^{\alpha}(s)}}{s - m^2 + \rmi m \sum\limits_{\dop \alpha,
    d} \Gamma_d^{\dop \alpha}(s)} .
\label{tmatapprox}
\ee
Here $\sum\limits_{\dop \alpha, d} \Gamma_d^{\dop \alpha}(s)$ denotes
the total decay width of the resonance summed over all quantum numbers
$\dop \alpha$ and decay channels $d$. At first glance this expression
is very similar to the one obtained in the $K$-matrix approach for the
case of only a single resonance contribution (see Eqn.\ (\ref{kmat})):
\be 
T_{fi}^{\alpha} = \left ( \frac{K^{\alpha}}{1 - \rmi K^{\alpha}}
\right )_{fi} = \frac{- m \sqrt {\Gamma_f^{\alpha}(s)
    \Gamma_i^{\alpha}(s)}}{s - m^2 + \rmi m \sum\limits_{d}
  \Gamma_d^{\alpha}(s)} .
\label{kmatapprox}
\ee
Here $K^{\alpha}$ is the full $n \times n$ matrix. The difference to
(\ref{tmatapprox}) is that the sum in the denominator runs over the
possible decay channels only. If $K^{\alpha}$ contains contributions
from different resonances/diagrams than it is no longer possible to
write $T_{fi}^{\alpha}$ in the form (\ref{kmatapprox}). Additionally,
in the $T$-matrix approximation the background contributions are
purely real, whereas in the $K$-matrix formalism also the imaginary
parts of these amplitudes are generated.

Calculating the $T$-matrix with the use of (\ref{tmatapprox}) violates
unitarity, because all rescattering contributions to a reaction $i \to
f$ via some intermediate state $d \ne i, f$ are neglected. To have a
measure for this violation in a specific channel $\alpha$, it is useful
to look at the following quantity:
\be 
\Delta T^{\alpha} = \imag (T^{\alpha}) - {\abs {T^{\alpha}}}^2 ,
\label{tmaterr}
\ee
which should vanish if unitarity is fulfilled. Again $T^{\alpha}$
denotes the $n \times n$ matrix. One expects $\Delta T^{\alpha}$ to be
negligible for channels where a single resonance gives the dominant
contribution (e.g. $D_{13}$ and $P_{33}$ in $\pi N$-scattering), since
there the expressions (\ref{tmatapprox}) and (\ref{kmatapprox}) agree
very well. This can be seen from the lower panel of Fig.\
\ref{unitarpp}. There the imaginary part of the $D_{13}$-partial wave
and $\Delta D_{13}$ are shown for a calculation employing the
$T$-matrix approximation. $\Delta D_{13}$ is small over the whole
energy range and vanishes on the $\nres D13{1520}$ mass. We can
further notice that the fit to the KA84-PWA is better than in the
$K$-matrix formalism (comp. Fig.\ \ref{ppi12KA84}). This is due to the
fact that here we do not have contributions to the imaginary part from
the background terms. Thus the real and imaginary parts of $T$ are
'decoupled' and can be fitted rather independently.

The situation is totally different in the $P_{13}$-partial wave (Fig.\
\ref{unitarpp}, upper panel). Here no satisfactory fit to the data can
be found. Especially at energies around 1.5 GeV we find additional
structure when using the approximation (\ref{tmatapprox}) that is
neither present in the data nor in the $K$-matrix results (Fig.\
\ref{ppi12KA84} and \ref{ppi12SM95}). This structure is due to the
contributions of the $\nres D13{1520}$ to $P_{13}$. As already
discussed in Sec. \ref{rescouplings}, the spin-$\dreih$-resonances have
offshell contributions to various channels that can be adjusted using
the $z$-parameters. In other words, the partial widths
$\Gamma_d^{\alpha}(s)$ are in general not equal to zero for channels
with quantum numbers that differ from those of the resonance
$\alpha_R$. Only on the resonance position we have
\be \Gamma_d^{\alpha \ne \alpha_R}(s = m_R^2) = 0. \ee
In the $T$-matrix approximation (\ref{tmatapprox}) the width in the
propagator is taken to be $\sum_{\dop \alpha, d} \Gamma_d^{\dop
  \alpha}(s)$ for all channels (Eqn.\ \ref{tmatapprox}) and does not
vanish on the resonance. Since the offshell contributions of the
spin-$\dreih$-resonances to channels $\alpha \ne \alpha_R$ always
change sign on the resonance position, the resulting amplitudes
develops structure as a function of $s$. For the $K$-matrix ansatz
(\ref{kmatapprox}) this is not the case because in these channels both
numerator and denominator go through zero on the resonance mass and
the amplitude remains smooth. The artificial structures in the
$T$-matrix approximations, introduced by spin-$\dreih$-resonances,
have already been observed in other effective Lagrangian calculations
\cite{fm97}. From this we conclude that a meaningful fit to {\it all}
partial waves can only be done in the $K$-matrix approximation. In the
fits using the $T$-matrix approach this shows up as an increased
$\chi^2$ value, which is in the order of $\approx$ 15 for the use of
the KA84-PWA (as compared to 2 in the $K$-matrix calculation).

As already mentioned, rescattering contributions with $d \ne i, f$ are
neglected in the $T$-matrix approach. To illustrate the importance of
these contributions, we show the real part of the $S_{11}$-partial wave
for $\pi N \to \eta N$ in Fig.\ \ref{unitarpe}. The $K$-matrix
calculation both with and without the $\nres S11{1650}$ resonance are
compared to the $T$-matrix result. In the $K$-matrix approach the
$\nres S11{1650}$ has a strong influence even though it's $\eta N$
coupling is zero. In the $T$-matrix calculation this is not the case
so that there all other couplings need to be adjusted to simulate the
influence of the $\nres S11{1650}$. Especially the nonresonant
parameters can therefore be viewed as {\it effective} couplings only.

%
%

\section{Summary and conclusion}

In this paper we have presented a unitary description for meson
nucleon scattering based on the $K$-matrix approximation. The
potential is determined by contributions of the nucleon, $I = \einh,
\dreih$-resonances and meson-exchanges in the $t$-channel. Effective
Lagrangians are used to describe the couplings and form factors are
taken into account at the hadronic vertices.

Within this approach we are able to describe all data of the reactions
$\pi N \to \pi N$, $\pi N \to \pi\pi N$, $\pi^- p \to \eta n$ and
$\pi^- p \to K^0 \Lambda$ by the same set of parameters. The explicit
inclusion of the $\eta N$- and $K \Lambda$-final state enables us to
extract decays of the resonances more reliably than by just using the
$\pi N$-inelasticities. Our couplings and branching ratios are in good
agreement with the values found in other calculations for the strongly
excited resonances and show only minor deviations for the weakly
coupling states. The pole positions and residues have been estimated
and have been found to be also in good agreement with other results.
Further work is clearly needed to continue the $T$-matrix analytically 
into the complex energy plane to locate the resonance poles more
reliably. Nevertheless, we have shown that an effective Lagrangian
ansatz is capable of describing the coupled channel dynamics
adequately.

To estimate the systematic error in the determination of resonance
parameters, we have performed 6 different analyses: i) the $\pi
N$-PWA's KA84 and SM95 were used as an input, and ii) the fits were
done with three different combinations of form factors. We have found
that we can reproduce the KA84-data somewhat better than the
SM95-solution, mainly because the latter is an energy-independent
solution and exhibits a larger scattering than the KA84-PWA.

One of the most important features of our analysis is that the
nonresonant background is consistently generated from Feynman diagrams
and thus the number of free parameters is reduced considerably.
Furthermore, the background is not independently determined for each
partial wave. In the fits this leads to a smaller $\rho NN$-coupling
than usual. In order to circumvent this problem one would have to
modify the $\rho$-contribution to obtain a Regge-like behavior. The
smaller coupling in turn influences the masses and couplings of the
resonances, especially for the $\nres P33{1232}$ and the $\nres
D13{1520}$. Except for the $\rho NN$-coupling, the other nucleon-meson
couplings we find are reasonable and stable between the different
fits.

A point of special interest is the $\nres S11{1535}$, due to its large
$\eta N$-decay width. Here the extraction of accurate couplings would
be very helpful. Unfortunately, we find a large systematic uncertainty
coming from the form factors used. Especially the mass of the
resonance is not well constrained by the available $\pi^- p \to \eta
n$ data. Since all fits and models describe the available data (see
Fig.\ \ref{pekcomp}), only new measurements would help to clarify the
situation. A search for a resonance pole of the $\nres S11{1535}$
within our approach would be very valuable to help to understand the
nature of this resonance.

The $z$-parameters of the spin-$\dreih$-resonances have been
investigated systematically. For the $I = \einh$ case, these
parameters exhibit large systematic errors and cannot be determined
very accurately because the large number of resonances and open
channels smear out the offshell-contributions. Accordingly, the fits
are more stable for the $I = \dreih$-resonances. The values for
$z_{\pi}$ of the $\Delta$ that we find are in good agreement with
previous determinations.

Our results indicate that a better fit to the $\pi N$-data could be
possible with the use of form factors that are not symmetric around
the resonance position. Especially for the spin-$\dreih$ cases a
significant improvement might be achieved with a functional form
closer to the usual dipoles. This needs to be investigated in more
detail.

The accuracy of the extracted parameters is limited mostly because of
the poor quality of the $\eta N$ and $K \Lambda$ data. From these the
corresponding partial widths cannot be determined to better than
$\approx$ 10-20 MeV. Also the resonance positions carry the same
error. New measurements could improve the situation, but at the same
time a better understanding of the differences between the $\pi N$-
and the $\pi\pi N$-PWA's is needed.

As already pointed out, another possible source of information is the
photoproduction of mesons. Especially for the case of
$\eta$-production high-quality data are available from recent
measurements \cite{k95}. A combined analysis of the hadronic and
electromagnetic reaction channels might put stricter limits on the
resonance parameters.

\section{Acknowledgments}

One of the authors (U.M.) thanks the Institute for Nuclear Theory at
the University of Washington for its hospitality and the U.S.
Department of Energy for partial support during completion of this
work.

%
%

\appendix

\section{Extraction of partial wave amplitudes}
\label{decompose}

In this appendix we derive the relations between the Feynman matrix
elements and the partial-wave decomposition of the meson-nucleon
scattering. For the $\pi N$-case these relations are well known and
given in standard textbooks \cite{gw64,ew88}. We use the metric of
Bjorken and Drell in the following \cite{bd66}. $p, \dop p, q$ and
$\dop q$ denote the four-momenta of the initial and final hadron and
the initial and final meson. ${\mathrm p}, {\mathrm \dop p}, {\mathrm
  q}$ and ${\mathrm \dop q}$ are the corresponding absolute values of
the three-momenta \mathbold p, \mathbold {\dop p}, \mathbold q and
\mathbold {\dop q}.

\subsection{Mesons of equal parity}

If both initial and final meson have the same parity, the Feynman
amplitude for meson nucleon scattering is given by ($Q = (q + \dop
q$)/2 is the average of the meson momenta):
\be {\cal M}_{fi} = \bar u (\dop p, \dop s) \left ( A + B \bsl Q
\right ) u (p, s) .
\label{amesonpos}
\ee
In terms of Pauli spinors the scattering amplitude, on the other hand,
can by written as \cite{gw64}:
\be 
F = \chi_{\dop s}^{\dagger} ( \tilde A + \tilde B \, \mathbold
{\sigma \cdot \hatdop p} \; \mathbold {\sigma \cdot \hat p} ) \chi_{s}
, 
\quad \mathbold {\hat p} = \frac{\mathbold p}{{\mathrm p}}, \qquad
\mathbold {\hatdop p} = \frac{\mathbold {\dop p}}{{\mathrm \dop p}} ,
\ee
with the well known decomposition:
\bea F &=& \frac{1}{\sqrt {\mathrm q \dop q}} \, \sumlUNE \, [ l
T_{l-} + (l+1) T_{l+} ] P_l -
\rmi \mathbold {\sigma \cdot} (\mathbold {\hatdop p \times \hat p}) 
[ T_{l+} - T_{l-} ] \dop {P_l} \nonumber \\
T_{l\pm} &=& \frac{1}{2} \intm1p1 d \cos \theta \tilde A P_l(\cos \theta) -
\tilde B P_{l\pm}(\cos \theta) .
\label{apartialpos}
\eea
The relation between the amplitudes $A, B$ and their counterparts
$\tilde A, \tilde B$ can be derived by inserting the explicit
representation of the spinors and $\gamma$-matrices in
(\ref{amesonpos}). Taking into account the different masses of the
initial and final mesons leads to:
\bea \tilde A &=&
\frac{\sqrt {(\dop E + \dop m)(E + m)}}{8 \pi \sqrt s} 
(A + B (\sqrt s - \bar m)) \nonumber \\
\tilde B &=& -
\frac{\sqrt {(\dop E - \dop m)(E - m)}}{8 \pi \sqrt s} 
(A - B (\sqrt s + \bar m)) \nonumber \\
\bar m &=& \frac{\dop m + m}{2} .
\label{amesonpospart}
\eea

\subsection{Mesons with different parity}

For scattering of mesons with different parity the starting point is
\bea {\cal M}_{fi} &=& \bar u (\dop p, \dop s) \gamma_5 \left ( A + B
  \bsl Q \right )
u (p,s) \nonumber \\
F &=& \chi_{\dop s}^{\dagger} ( \tilde A \; \mathbold {\sigma \cdot
  \hatdop p} + \tilde B \; \mathbold {\sigma \cdot \hat p} ) \chi_{s}
,
\label{amesonneg}
\eea
with the decomposition
\bea F &=& \frac{1}{\sqrt {\mathrm q \dop q}} \, \sumlUNE \, [ l
T_{l-} + (l+1) T_{l+} ] P_l +
\rmi \mathbold {\sigma \cdot} (\mathbold {\hat p \times \hatdop p}) 
[ T_{l+} - T_{l-} ] \dop {P_l} \nonumber \\
T_{l\pm} &=& \frac{1}{2} \intm1p1 d \cos \theta \tilde A P_l(\cos \theta) +
\tilde B P_{l\pm}(\cos \theta) .
\label{apartialneg}
\eea
An analogous calculation as in the equal-parity case yields the
relations between $A, B$ and $\tilde A, \tilde B$:
\bea \tilde A &=& -
\frac{\sqrt {(\dop E - \dop m)(E + m)}}{8 \pi \sqrt s} 
(A + B (\sqrt s + \partial m)) \nonumber \\
\tilde B &=&
\frac{\sqrt {(\dop E + \dop m)(E - m)}}{8 \pi \sqrt s} 
(A - B (\sqrt s - \partial m)) \nonumber \\
\partial m &=& \frac{\dop m - m}{2} .
\label{amesonnegpart}
\eea

\subsection{Isospin decomposition}

For the $I=1$ mesons $\pi$ and $\zeta$ we start from the standard
projection operators \cite{ew88}
\bea
P_{\einh} &=& \frac{1}{3}(1 - \mathbold {t \cdot \tau}) \nonumber \\
P_{\dreih} &=& \frac{1}{3}(2 + \mathbold {t \cdot \tau}) , \eea
with the matrix elements (a, b = $\pi, \zeta$)
\bea
\langle b_j | P_{\einh} | a_i \rangle &=& 
\frac{1}{3} \tau_j \tau_i \nonumber \\
\langle b_j | P_{\dreih} | a_i \rangle &=& \delta_{ji} - \frac{1}{3}
\tau_j \tau_i \eea
in a cartesian basis. With the help of this all possible reactions can
be written as:
\bea \langle b_j N | T_{\Iba} | a_i N \rangle &=& \frac{1}{3} \tau_j
\tau_i T_{\Iba}^{\einh} +
(\delta_{ji} - \frac{1}{3} \tau_j \tau_i) T_{\Iba}^{\dreih} , 
\label{decompi1} \\
\mbox{explicitly :} \nonumber \\
\langle b^+ p | a^+ p \rangle &=& T^{\dreih}_{\Iba} \nonumber \\
\langle b^- p | a^- p \rangle &=& \frac{1}{3}
(T^{\dreih}_{\Iba} + 2 T^{\einh}_{\Iba}) \nonumber \\
\langle b^- p | a^0 n \rangle &=& \frac{\sqrt 2}{3}
(T^{\dreih}_{\Iba} - T^{\einh}_{\Iba}) \nonumber \\
&\cdots& , \nonumber \eea with the factors being the corresponding
Clebsch-Gordan coefficients.

For the pure $I=\einh$-reactions involving $\pi$ and $\zeta$ the
projector is usually taken to be $P_{\einh} = \mathbold \tau$
\cite{ew88}. This choice has the disadvantage that it does not agree
with the Clebsch-Gordan coefficients for the different reactions
channels. Therefore we here choose (a = $\pi, \zeta$, b = $\eta, k$):
\bea
\langle b | P_{\einh} | a_i \rangle &=& \frac{-1}{\sqrt 3} \tau_i \nonumber \\
\langle b | T_{\Iba} | a_i \rangle &=& \frac{-1}{\sqrt 3} \tau_i
T^{\einh}_{\Iba}.
\label{decompi0}
\eea
This has no influence on the calculated quantities, since in the end,
we convert our amplitudes to the normal convention.

\section{Observables}
\label{observables}

For completeness we also list the formulas need for calculating the
different observables from the partial waves. $P_l$ and $\dop P_l$
denote the Legendre polynomials and their derivatives.

Total cross sections $\sigma$:
\be \sigma = \frac{4 \pi}{{\mathrm q}^2} \sum\limits_{l=0}^{l_{max}}
\left ( (l+1) {\abs {\tilde T_{l+}}}^2 + l {\abs {\tilde T_{l-}}}^2
\right ) , \ee
differential cross sections $\frac{d \sigma}{d \Omega}$ and
final-state polarizations $P$:
\bea f &=& \frac{1}{\mathrm q} \sum\limits_{l=0}^{l_{max}}
\left ( (l+1) \tilde T_{l+} + l \tilde T_{l-} \right ) P_l \nonumber \\
g &=& \frac{1}{\mathrm q} \sin \theta \sum\limits_{l=0}^{l_{max}}
\left ( \tilde T_{l+} - \tilde T_{l-} \right ) \dop P_l \nonumber \\
\frac{d \sigma}{d \Omega} &=& {\abs f}^2 + {\abs g}^2 , \quad \frac{d
  \sigma}{d \Omega} P = - 2 \imag (f^* g) . \eea
Here $\tilde T_{l\pm}$ denotes the partial wave amplitude for a
specific reaction. It is given as a sum over the contributing
isospin-channels:
\be \tilde T_{l\pm} = \sum\limits_I p^I T_{l\pm}^I. \ee
The factors $p^I$ can be determined from (\ref{decompi1}) and
(\ref{decompi0}).

Inelastic cross section $\sigma_{inel}$:
\be \sigma_{inel} = \frac{4 \pi}{{\mathrm q}^2} \left ( \imag
  (T^{\alpha}_{\pi N}) - {\abs {T^{\alpha}_{\pi N}}}^2 \right ) \ee

\section{Coupling constants and decay widths}
\label{ccanddw}

In this appendix we list the formulas for the decay widths as
calculated from the Lagrangians given in Sec. \ref{rescouplings}. Here
${\mathrm p}$ denotes the three-momentum of the meson and nucleon,
$E_N$ and $E_{\varphi}$ the nucleon and meson energy, respectively:
\bea
{\mathrm p} &=& \frac{\sqrt{(s - (m_N + m_{\varphi})^2)
(s - (m_N - m_{\varphi})^2)}}{2 \sqrt s} \nonumber\\
E_N &=& \sqrt{{\mathrm p}^2 + m_N^2} , \qquad E_{\varphi} =
\sqrt{{\mathrm p}^2 + m_{\varphi}^2} . \eea

For spin-$\einh$-resonances we have:
\bea
\mbox {PS-coupling} &:& \nonumber \\
\Gamma_{\pm} &=& {\mathrm ISO} \; \frac{g_{\varphi NR}^2}{4 \pi} \; 
{\mathrm p} \; \frac{E_N \mp m_N}{\sqrt s} \nonumber \\
\mbox {PV-coupling} &:& \nonumber \\
\Gamma_{\pm} &=& {\mathrm ISO} \; \frac{g_{\varphi NR}^2}{4 \pi (m_R
  \pm m_N)^2} \; {\mathrm p} \; \frac{2 E_{\varphi} (E_N E_{\varphi} +
  {\mathrm p}^2) - m_{\varphi}^2 (E_N \pm m_N)}{\sqrt s} . \eea
The upper sign corresponds to decays of resonances into mesons with
opposite parity (e.g. $\nres P11{1440} \to \pi N$), the lower sign
holds if both have the same parity (e.g. $\nres S11{1535} \to \pi N$).
$\mathrm ISO$ is the isospin factor, it is equal to 3 for decays into
mesons with isospin one, 1 otherwise.

Spin-$\dreih$-resonances:
\be \Gamma_{\pm} = \frac{g_{\varphi NR}^2}{12 \pi m_{\pi}^2} \;
{\mathrm p}^3 \; \frac{E_N \pm m_N}{\sqrt s} . \ee
Again, the upper sign is used if resonance and meson are of opposite
parity.

%
%

\newpage

%
%



\pagestyle{empty}

\setcounter{table}{0} \setcounter{figure}{0}

\begin{table}[ht]
\begin{center}
  \renewcommand{\arraystretch}{1.2}
\begin{tabular}{c|r|c|c|c|r|c|c|c|c|c|c|c}
               & $M$ & S        & I        & P & $\Gamma_{tot}$ &
$\Gamma_{\pi \pi}$ & $\Gamma_{\pi \eta}$ & $\Gamma_{\pi k}$ & $\Gamma_{\pi \gamma}$ & $\Gamma_{\eta \gamma}$ & $\Gamma_{k \gamma}$ & $\Gamma_{\gamma \gamma}$ \\
               &[GeV]&          &          &   & [MeV]          &
[\%]               & [\%]                & [\%]             & [\%]                  &
[\%]                   & [\%]                & [\%] \\
\hline\hline
$\pi$     & 0.139 & 1       & 1 & -- & 7.85$^a$ & 0   & 0   & 0   & 0   & 0   & 0   & 99 \\
$\zeta^c$ & 0.278 & 1       & 1 & +  & --       & --  & --  & --  & --  & --  & --  & -- \\
$\eta$    & 0.548 & 0       & 0 & -- & 1.2$^b$  & 0   & 0   & 0   & 0   & 0   & 0   & 39 \\
$K^c$     & 0.498 & 0       & $\einh$ & -- & --       & --  & --  & --  & --  & --  & --  & -- \\
\hline\hline
$\rho$    & 0.769 & 1       & 1 & -- & .151     & 100 & 0   & 0   & .05 & .04 & 0   & 0 \\
$a_0$     & 0.983 & 0       & 1 & +  & .200     & 0   & 100 & 0   & 0   & 0   & 0   & 0 \\
$K^*$     & 0.892 & 1       & $\einh$ & -- & .050     & 0   & 0   & 100 & 0   & 0   & .1  & 0 \\
\end{tabular} 
\renewcommand{\arraystretch}{1.0}
\end{center}
\caption{Masses and widths of the mesons included. $\pi, \zeta, \eta$
  and $K$ are the asymptotic states. $^a$: Width in $eV$, $^b$: Width
  in $keV$, $^c$: no decays were taken into account. Data as given by
  the Particle Data Group \protect\cite{pdg96}.}
\label{mesdata}
\end{table}

\begin{table}[!ht]
\begin{center}
\begin{tabular}{l|c|c|c|c|c|c}
        & $\chi^2$ & $\chi^2/{\rm DF}$ & $\chi^2_{\pi}/{\rm DF}$ & 
$\chi^2_{\pi\pi}/{\rm DF}$ & $\chi^2_{\eta}/{\rm DF}$ & 
$\chi^2_{K}/{\rm DF}$ \\
\hline\hline
KA84-pp & 4196     & 2.84 & 2.50 & 6.52 & 1.42 & 3.14 \\
KA84-ee & 4616     & 3.13 & 2.99 & 5.59 & 1.58 & 3.52 \\
KA84-pt & 4067     & 2.76 & 2.41 & 5.70 & 1.50 & 3.39 \\
\hline
SM95-pp & 4720     & 3.62 & 3.78 & 6.27 & 1.49 & 3.31 \\
SM95-ee & 4871     & 3.74 & 4.11 & 5.60 & 1.61 & 3.28 \\
SM95-pt & 4574     & 3.52 & 3.69 & 5.67 & 1.64 & 3.22 \\
\end{tabular} 
\end{center}
\caption{$\chi^2$-values for the different fits. $\chi^2/{\rm DF}$
  gives the $\chi^2$ per datapoint. Also the $\chi^2/{\rm DF}$-values 
for the different reaction channels are given separately.}
\label{chi2comp}
\end{table}

\begin{table}[!ht]
\begin{center}
\begin{tabular}{c|c|r|r|r}
\multicolumn{2}{c|}{} & \multicolumn{1}{c|}{KA84-pp} & 
        \multicolumn{1}{c|}{SM95-pt} & \multicolumn{1}{c}{Others} \\
\multicolumn{2}{c|}{} & \multicolumn{1}{c|}{[fm]} & 
        \multicolumn{1}{c|}{[fm]} & \multicolumn{1}{c}{[fm]} \\
\hline\hline
 & $a_1$ &  0.180              &  0.168              &  0.247$^a$, 0.246$^b$, 0.252$^c$ \\
 & $r_1$ & -2.430              & -3.062              & \multicolumn{1}{c}{--} \\
\raisebox{1.5ex}[-1.5ex]{$\pi N$}
 & $a_3$ & -0.114              & -0.142              & -0.144$^a$, -0.130$^b$, -0.143$^c$ \\
 & $r_3$ &  13.300             &  7.668              & \multicolumn{1}{c}{--} \\
\hline
 &       &  0.487 + \rmi 0.171 &  0.577 + \rmi 0.216 &  0.51 + \rmi 0.21$^d$ \\
 & $a_1$ &                     &                     &  0.717(30) + \rmi 0.263(25)$^e$ \\
\raisebox{1.5ex}[-1.5ex]{$\eta N$}
 &       &                     &                     &  0.751(43) + \rmi 0.274(28)$^f$\\
 & $r_1$ & -6.060 - \rmi 0.177 & -2.807 - \rmi 0.057 & -1.496(134) - \rmi 0.237(37)$^f$\\
\hline
 & $a_1$ &  0.065 + \rmi 0.040 &  0.048 + \rmi 0.030 & \multicolumn{1}{c}{--} \\
\raisebox{1.5ex}[-1.5ex]{$K \Lambda$}
 & $r_1$ & -15.930 - \rmi 8.252 & -24.324 - \rmi 13.853 & \multicolumn{1}{c}{--} \\
\end{tabular} 
\end{center}
\caption{$\pi N$-, $\eta N$- and $K \Lambda$-scattering lengths as
  obtained in the fits in comparison with the results of other works.
  $^a$: \protect\cite{ew88}, $^b$: \protect\cite{sm95}, $^c$:
  \protect\cite{ka84}, $^d$: \protect\cite{sau96}, $^e$:
  \protect\cite{bdssnl97}, $^f$: \protect\cite{gw97}. Number in
  brackets indicate the error in the last digits.}
\label{parmcomp}
\end{table}

\begin{table}[ht]
\begin{center}
\begin{tabular}{c|c|r|c|r|c|r|c|r}
       & \multicolumn{4}{c|}{KA84} & \multicolumn{4}{c}{SM95} \\
\cline{2-9}
       & $g$                      & Value  & $\kappa$                 & Value  &
         $g$                      & Value  & $\kappa$                 & Value  \\
\hline\hline
$\pi$  & $g_{\pi NN}$             & 13.05  & -- & -- & 
         $g_{\pi NN}$             & 13.05  & -- & -- \\
       &                          & 13.06  & -- & -- &  
                                  & 13.04  & -- & -- \\
       &                          & 13.05  & -- & -- &
                                  & 13.05  & -- & -- \\
\hline
$\eta$ & $g_{\eta NN}$            & 1.08   & -- & -- & 
         $g_{\eta NN}$            & 1.33   & -- & -- \\
       &                          & 2.39   & -- & -- &
                                  & 0.18   & -- & -- \\
       &                          & 1.86   & -- & -- &
                                  & 1.13   & -- & -- \\
\hline
$K$    & $g_{K N \Lambda}$        & -6.56  & -- & -- &
         $g_{K N \Lambda}$        & -6.36  & -- & -- \\
       &                          & -6.41  & -- & -- & 
                                  & -6.10  & -- & -- \\
       &                          & -6.06  & -- & -- & 
                                  & -6.12  & -- & -- \\
\hline
$\rho$ & $g_{\rho NN}$            & 3.22   & $\kappa_{\rho NN}$       & 2.14   &  
         $g_{\rho NN}$            & 3.37   & $\kappa_{\rho NN}$       & 1.99   \\
       &                          & 3.38   &                          & 2.34   & 
                                  & 3.53   &                          & 2.35   \\
       &                          & 2.11   &                          & 2.65   & 
                                  & 2.35   &                          & 2.26   \\
\hline
$a_0$  & $g_{a_0 NN}$             & 1.57   & -- & -- & 
         $g_{a_0 NN}$             & 0.68   & -- & -- \\
       &                          & 3.33   & -- & -- & 
                                  & 2.55   & -- & -- \\
       &                          & 0.93   & -- & -- & 
                                  & 0.18   & -- & -- \\
\hline
$K^*$  & $g_{K^* N \Lambda}$      & -21.65 & $\kappa_{K^* N \Lambda}$ & -0.43  & 
         $g_{K^* N \Lambda}$      & -21.58 & $\kappa_{K^* N \Lambda}$ & -0.43  \\
       &                          & -21.99 &                          & -0.44  & 
                                  & -23.23 &                          & -0.43  \\
       &                          & -5.90  &                          & -0.44  & 
                                  & -6.52  &                          & -0.43  \\
\end{tabular} 
\end{center}
\caption{Couplings of the mesons to the nucleon as obtained in the
  fits. In the first columns we list the results of the fits KA84-pp, 
KA84-ee and KA84-pt, while in the other we give SM95-pp, SM95-ee and SM95-pt.}
\label{mesparmKA84SM95}
\end{table}

\begin{table}[!ht]
\begin{center}
\begin{tabular}{c|c|r|r|r|r}
                         & $q^2$   & KA84            & SM95            & 
        Others & SU(3) \\
\hline\hline
$g_{\pi NN}$             & $m^2$   & 13.05           & 13.05           & 
        13.14$^a$, 13.41$^b$ & 13.3 \\
                         & $q_s^2$ & 12.56 - 12.69   & 12.62 - 12.70   & 
        -- & -- \\
\hline
$g_{\eta NN}$            & $m^2$   & 1.08 - 2.39     & 0.18 - 1.33     & 
        4.1 - 6.3$^c$, 4.19$^d$ & 2.2 - 5.9 \\
                         & $q_s^2$ & 0.57 - 1.26     & 0.10 - 0.77     & 
        -- & -- \\
\hline
$g_{K N \Lambda}$        & $m^2$   & -(6.06 - 6.56)  & -(6.10 - 6.36)  & 
        -14.78$^e$, -10.96$^f$ & -(10.3 - 16.7) \\
                         & $q_s^2$ & -(2.19 - 2.62)  & -(2.17 - 2.84)  & 
        -5.35$^f$ &             -- \\
\hline\hline
$g_{\rho NN}$            & $m^2$   & 2.11 - 3.38     & 2.35 - 3.37     & 
        3.14$^b$, 2.63$^g$ & 2.66 \\
                         & $q_t^2$ & 2.07 - 2.11     & 1.98 - 2.35     & 
        2.67$^b$ & -- \\
$\kappa_{\rho NN}$       & $q_t^2$ & 2.14 - 2.65     & 1.99 - 2.35     & 
        2.25$^b$, 3.71$^g$ & 3.71 \\
\hline
$g_{a_0 NN}$             & $m^2$   & 0.75 - 3.33     &  0.18 - 2.55    & 
        -- & -- \\
                         & $q_t^2$ & 0.53 - 0.75     &  0.18 - 0.30    & 
        -- & -- \\
\hline
$g_{K^* N \Lambda}$      & $m^2$   & -(5.90 - 21.99) & -(6.52 - 23.23) & 
        -(18.87 - 21.36)$^e$, -9.39$^f$ & -(3.69 - 5.53) \\
                         & $q_t^2$ & -(4.44 - 7.94)  & -(3.57 - 7.53)  & 
        -- & -- \\
$\kappa_{K^* N \Lambda}$ & $q_t^2$ & -0.44           & -0.43           & 
        -(0.43 - 0.72)$^e$, 0.59$^f$ & (1.48 - 2.23) \\
\end{tabular} 
\end{center}
\caption{Effective couplings ($g \cdot F(q^2)$) to the nucleon on the
  mass shell and at threshold. In the first two columns we give the
  lower and upper values from Table \protect\ref{mesparmKA84SM95}. The
  quoted values are taken from $^a$: \protect\cite{sm95}, $^b$: 
\protect\cite{pj91}, $^c$: \protect\cite{bmz95}, $^d$: 
\protect\cite{tbk94}, $^e$: \protect\cite{as90}, $^f$: 
\protect\cite{bmk97}, $^g$: \protect\cite{hp75}. The SU(3)-predictions
use the given values for $g_{\pi NN}$, $g_{\rho NN}$ and $\kappa_{\rho
  NN}$ and include symmetry-breaking on the level of 20\% 
\protect\cite{t66}.}
\label{mescoupl}
\end{table}

\begin{table}[!ht]
\begin{center}
\begin{tabular}{c|r|r|r|r|r|r|r|r|r|r}
                 & $M$ & $\Gamma_{tot}$ &
\multicolumn{2}{c|}{$\Gamma_{\pi N}$}  & \multicolumn{2}{c|}{$\Gamma_{\zeta N}$} & \multicolumn{2}{c|}{$\Gamma_{\eta N}$} & \multicolumn{2}{c}{$\Gamma_{K \Lambda}$} \\
$L_{2I,2S}$       &[GeV]  & [MeV]& [MeV]&   \% & [MeV]&   \% & [MeV]&   \% & [MeV]&   \% \\
\hline\hline
$\nres S11{1535}$ & 1.550 &  240 &  120 &   50 &   -- &   -- &   -- &   -- &   -- &   -- \\
                  & 1.526 &  120 &   46 &   38 &   -- &   -- &   -- &   -- &   -- &   -- \\
                  & 1.535 &   66 &   20 &   31 &   -- &   -- &   -- &   -- &   -- &   -- \\
                  & 1.534 &  151 &   77 &   51 &   10 &    5 &   66 &   43 &    0 &    0 \\
                  & 1.553 &  182 &   84 &   46 &    7 &    4 &   91 &   50 &   -- &   -- \\
                  & 1.547 &  162 &   66 &   41 &    6 &    4 &   89 &   55 &   -- &   -- \\
                  & 1.534 &  125 &   53 &   42 &   19 &   15 &   54 &   43 &   -- &   -- \\[0.5ex]
$\nres S11{1650}$ & 1.650 &  150 &   98 &   65 &   -- &   -- &   -- &   -- &   -- &   -- \\
                  & 1.670 &  180 &  110 &   61 &   -- &   -- &   -- &   -- &   -- &   -- \\
                  & 1.667 &   90 &   90 &  100 &   -- &   -- &   -- &   -- &   -- &   -- \\
                  & 1.659 &  173 &  154 &   89 &   13 &    8 &    6 &    3 &    0 &    0 \\
                  & 1.652 &  202 &  160 &   79 &   16 &    8 &   26 &   13 &   -- &   -- \\
                  & 1.695 &  293 &  226 &   77 &   67 &   23 &   -- &   -- &   -- &   -- \\
                  & 1.690 &  229 &  149 &   65 &   23 &   10 &   57 &   25 &   -- &   -- \\[0.5ex]
$\nres S11{2090}$ & 2.180 &  350 &   63 &   18 &   -- &   -- &   -- &   -- &   -- &   -- \\
                  & 1.880 &   95 &    9 &    9 &   -- &   -- &   -- &   -- &   -- &   -- \\
                  & 1.712 &  184 &   70 &   38 &   -- &   -- &   -- &   -- &   -- &   -- \\
                  & 1.928 &  414 &   43 &   10 &  369 &   90 &    2 &    0 &    0 &    0 \\
                  & 1.812 &  405 &  130 &   32 &  186 &   46 &   89 &   22 &   -- &   -- \\
\hline
$\nres P11{1440}$ & 1.440 &  340 &  231 &   68 &   -- &   -- &   -- &   -- &   -- &   -- \\
                  & 1.410 &  135 &   69 &   51 &   -- &   -- &   -- &   -- &   -- &   -- \\
                  & 1.467 &  440 &  299 &   68 &   -- &   -- &   -- &   -- &   -- &   -- \\
                  & 1.462 &  391 &  270 &   69 &  121 &   31 &    0 &    0 &    0 &    0 \\
                  & 1.439 &  437 &  271 &   62 &  166 &   38 &    0 &    0 &   -- &   -- \\[0.5ex]
$\nres P11{1710}$ & 1.700 &   90 &   18 &   20 &   -- &   -- &   -- &   -- &   -- &   -- \\
                  & 1.723 &  120 &   14 &   12 &   -- &   -- &   -- &   -- &   -- &   -- \\
                  &    -- &   -- &   -- &   -- &   -- &   -- &   -- &   -- &   -- &   -- \\
                  & 1.717 &  478 &   45 &    9 &  249 &   52 &   10 &    2 &  175 &   37 \\
                  & 1.729 &  180 &   40 &   22 &  130 &   72 &   11 &    6 &   -- &   -- \\
\end{tabular} 
\end{center}
\caption{Resonance masses and couplings ($I = \einh$, $S = \einh$) as 
obtained in other models. For each resonance we list in lines one to
five the values of Cutkosky et al. \protect\cite{cfhk79}, H\"ohler et
al. \protect\cite{ka84}, Arndt et al. \protect\cite{sm95}, Manley et
al. \protect\cite{ms92} and Batinic et al. \protect\cite{bdssnl97}.
 Furthermore the $S_{11}$ parameters from \protect\cite{sau96} (line
 5) and  \protect\cite{dvl97} ($K$-matrix result, line 6) are given. 
Only in \protect\cite{ms92} a $K \Lambda$-decay was included.}
\label{rescoupl1212comp}
\end{table}

\begin{table}[!ht]
\begin{center}
\begin{tabular}{c|r|r|r|r|r|r|r|r|r|r}
                 & $M$ & $\Gamma_{tot}$ &
\multicolumn{2}{c|}{$\Gamma_{\pi N}$}  & \multicolumn{2}{c|}{$\Gamma_{\zeta N}$} & \multicolumn{2}{c|}{$\Gamma_{\eta N}$} & \multicolumn{2}{c}{$\Gamma_{K \Lambda}$} \\
$L_{2I,2S}$       &[GeV]  & [MeV]& [MeV]&   \% & [MeV]&   \% & [MeV]&   \% & [MeV]&   \% \\
\hline\hline
$\nres P13{1720}$ & 1.700 &  125 &   13 &   10 &   -- &   -- &   -- &   -- &   -- &   -- \\
                  & 1.710 &  190 &   27 &   14 &   -- &   -- &   -- &   -- &   -- &   -- \\
                  & 1.820 &  354 &   57 &   16 &   -- &   -- &   -- &   -- &   -- &   -- \\
                  & 1.717 &  383 &   50 &   13 &  333 &   87 &    0 &    0 &    0 &    0 \\
                  & 1.720 &  244 &   44 &   18 &  200 &   82 &    1 &  0.4 &   -- &   -- \\
\hline
$\nres D13{1520}$ & 1.525 &  120 &   70 &   58 &   -- &   -- &   -- &   -- &   -- &   -- \\
                  & 1.519 &  114 &   62 &   54 &   -- &   -- &   -- &   -- &   -- &   -- \\
                  & 1.515 &  106 &   65 &   61 &   -- &   -- &   -- &   -- &   -- &   -- \\
                  & 1.524 &  124 &   73 &   59 &   51 &   41 &    0 &    0 &    0 &    0 \\
                  & 1.522 &  132 &   73 &   55 &   59 &   45 &    1 &  0.1 &   -- &   -- \\[0.5ex]
$\nres D13{1700}$ & 1.675 &   90 &   10 &   11 &   -- &   -- &   -- &   -- &   -- &   -- \\
                  & 1.731 &  110 &    9 &    8 &   -- &   -- &   -- &   -- &   -- &   -- \\
                  &    -- &   -- &   -- &   -- &   -- &   -- &   -- &   -- &   -- &   -- \\
                  & 1.737 &  249 &    0 &    1 &  241 &   98 &    5 &    2 &    0 &    0 \\
                  & 1.817 &  134 &   12 &    9 &  103 &   77 &   19 &   14 &   -- &   -- \\[0.5ex]
$\nres D13{2080}$ & 1.880 &  180 &   18 &   10 &   -- &   -- &   -- &   -- &   -- &   -- \\
                  & 2.081 &  265 &   16 &    6 &   -- &   -- &   -- &   -- &   -- &   -- \\
                  &    -- &   -- &   -- &   -- &   -- &   -- &   -- &   -- &   -- &   -- \\
                  & 1.804 &  447 &  104 &   23 &  224 &   50 &  119 &   27 &    0 &    0 \\
                  & 2.048 &  529 &   90 &   17 &  397 &   75 &   42 &    8 &   -- &   -- \\
\end{tabular} 
\end{center}
\caption{Same as Table \protect\ref{rescoupl1212comp}, but for the 
$I = \einh$-$S = \dreih$-resonances.}
\label{rescoupl1232comp}
\end{table}

\begin{table}[!ht]
\begin{center}
\begin{tabular}{c|r|r|r|r|r|r|r|r|r|r}
                 & $M$ & $\Gamma_{tot}$ &
\multicolumn{2}{c|}{$\Gamma_{\pi N}$}  & \multicolumn{2}{c|}{$\Gamma_{\zeta N}$} & \multicolumn{2}{c|}{$\Gamma_{\eta N}$} & \multicolumn{2}{c}{$\Gamma_{K \Lambda}$} \\
$L_{2I,2S}$       &[GeV]  & [MeV]& [MeV]&   \% & [MeV]&   \% & [MeV]&   \% & [MeV]&   \% \\
\hline\hline
$\nres S31{1620}$ & 1.620 &  140 &   35 &   25 &   -- &   -- &   -- &   -- &   -- &   -- \\
                  & 1.610 &  139 &   49 &   35 &   -- &   -- &   -- &   -- &   -- &   -- \\
                  & 1.617 &  108 &   31 &   29 &   -- &   -- &   -- &   -- &   -- &   -- \\
                  & 1.672 &  154 &   14 &    9 &  140 &   81 &   -- &   -- &   -- &   -- \\
\hline
$\nres P33{1232}$ & 1.232 &  120 &  120 &  100 &   -- &   -- &   -- &   -- &   -- &   -- \\
                  & 1.233 &  116 &  116 &  100 &   -- &   -- &   -- &   -- &   -- &   -- \\
                  & 1.233 &  114 &  114 &  100 &   -- &   -- &   -- &   -- &   -- &   -- \\
                  & 1.231 &  118 &  118 &  100 &    0 &    0 &   -- &   -- &   -- &   -- \\[0.5ex]
$\nres P33{1600}$ & 1.600 &  300 &   54 &   18 &   -- &   -- &   -- &   -- &   -- &   -- \\
                  & 1.522 &  220 &   46 &   21 &   -- &   -- &   -- &   -- &   -- &   -- \\
                  &    -- &   -- &   -- &   -- &   -- &   -- &   -- &   -- &   -- &   -- \\
                  & 1.706 &  430 &   53 &   12 &  377 &   87 &   -- &   -- &   -- &   -- \\
\hline
$\nres D33{1700}$ & 1.710 &  280 &   34 &   12 &   -- &   -- &   -- &   -- &   -- &   -- \\
                  & 1.680 &  230 &   46 &   20 &   -- &   -- &   -- &   -- &   -- &   -- \\
                  & 1.680 &  272 &   44 &   16 &   -- &   -- &   -- &   -- &   -- &   -- \\
                  & 1.762 &  599 &   81 &   14 &  518 &   86 &   -- &   -- &   -- &   -- \\
\end{tabular} 
\end{center}
\caption{Same as Table \protect\ref{rescoupl1212comp}, but for the 
$I = \dreih$-resonances. Given are the values of Cutkosky et al. 
\protect\cite{cfhk79}, H\"ohler et al. \protect\cite{ka84}, Arndt et
al. \protect\cite{sm95} and Manley and Saleski \protect\cite{ms92}.}
\label{rescoupl32comp}
\end{table}

\begin{table}[ht]
\begin{center}
\begin{tabular}{c|r|c|r|c|r|c|r}
            & Value &                   & Value &                    & Value &             & Value \\
            & [GeV] &                   & [GeV] &                    & [GeV] &             & [GeV] \\
\hline\hline
$\Lambda_N$ & 1.18  & $\Lambda_{\einh}$ & 1.59  & $\Lambda_{\dreih}$ & 1.04  & $\Lambda_t$ & 0.90  \\
            & 1.29  &                   & 1.82  &                    & 1.15  &             & 0.92  \\
            & 1.21  &                   & 1.72  &                    & 1.06  &             & 0.71  \\
\hline\hline
$\Lambda_N$ & 1.24  & $\Lambda_{\einh}$ & 1.36  & $\Lambda_{\dreih}$ & 1.06  & $\Lambda_t$ & 0.88  \\
            & 1.30  &                   & 1.71  &                    & 1.14  &             & 0.88  \\
            & 1.23  &                   & 1.24  &                    & 1.06  &             & 0.70  \\
\end{tabular} 
\end{center}
\caption{Values of the fitted cutoff-parameters $\Lambda$. KA84
  results are given in the first three rows (KA84-pp, KA84-ee and
  KA84-pt), below are the results using SM95 (SM95-pp, SM95-ee and 
SM95-pt).}
\label{cutoffsKA84SM95}
\end{table}

\begin{table}[ht]
\begin{center}
\begin{tabular}{c|r|r|r|r|r|r|r|r|r|r}
                 & $M$ & $\Gamma_{tot}$ &
\multicolumn{2}{c|}{$\Gamma_{\pi N}$}  & \multicolumn{2}{c|}{$\Gamma_{\zeta N}$} & \multicolumn{2}{c|}{$\Gamma_{\eta N}$} & \multicolumn{2}{c}{$\Gamma_{K \Lambda}$} \\
$L_{2I,2S}$       &[GeV]  & [MeV]&  [MeV] & \% &  [MeV] & \% &  [MeV] & \% &  [MeV] & \% \\
\hline\hline
$\nres S11{1535}$ & 1.534 &  180 &  71(+) &  39 &  14(+) &   8 &  95(+) &  53 &   0(+) &   0 \\
                  & 1.542 &  175 &  67(+) &  38 &   7(+) &   4 & 101(+) &  58 &   0(+) &   0 \\
                  & 1.542 &  198 &  74(+) &  38 &  10(+) &   5 & 113(+) &  57 &   0(+) &   0 \\[0.5ex]
$\nres S11{1650}$ & 1.694 &  212 & 157(+) &  74 &  38(+) &  18 &   1(-) &   0 &  16(+) &   8 \\
                  & 1.697 &  261 & 195(+) &  75 &  54(+) &  21 &   0(-) &   0 &  12(+) &   5 \\
                  & 1.701 &  278 & 205(+) &  74 &  61(+) &  22 &   1(-) &   0 &  11(+) &   4 \\
\hline
$\nres P11{1440}$ & 1.469 &  367 & 237(+) &  65 & 130(+) &  35 & 2.75$^a$ &   0 &   0(+) &   0 \\
                  & 1.476 &  412 & 269(+) &  65 & 143(+) &  35 & 4.22$^a$ &   0 &   0(+) &   0 \\
                  & 1.477 &  411 & 264(+) &  64 & 147(+) &  36 & 4.40$^a$ &   0 &   0(+) &   0 \\[0.5ex]
$\nres P11{1710}$ & 1.706 &  172 &   0(+) &   0 &  89(-) &  52 &  67(+) &  39 &  16(+) &   9 \\
                  & 1.696 &  123 &   0(+) &   0 &  71(-) &  58 &  19(+) &  15 &  33(+) &  27 \\
                  & 1.697 &  148 &   0(+) &   0 &  80(-) &  54 &  23(+) &  16 &  45(+) &  30 \\
\hline
$\nres P13{1720}$ & 1.790 &  384 &  84(+) &  22 & 259(+) &  67 &  36(+) &   9 &   5(+) &   1 \\
                  & 1.779 &  306 &  68(+) &  22 & 218(+) &  71 &  17(+) &   6 &   3(+) &   1 \\
                  & 1.803 &  480 & 107(+) &  22 & 324(+) &  68 &  44(+) &   9 &   5(+) &   1 \\
\hline
$\nres D13{1520}$ & 1.510 &  101 &  53(+) &  52 &  48(-) &  48 & 27$^b$(+) &   0 &   0(+) &   0 \\
                  & 1.510 &  100 &  54(+) &  54 &  46(-) &  46 & 44$^b$(+) &   0 &   0(+) &   0 \\
                  & 1.511 &   98 &  53(+) &  54 &  45(-) &  46 & 51$^b$(+) &   0 &   0(+) &   0 \\[0.5ex]
$\nres D13{1700}$ & 1.897 &  313 &  38(+) &  12 & 260(+) &  83 &  15(-) &   5 &   0(+) &   0 \\
                  & 1.888 &  303 &  41(+) &  14 & 259(+) &  85 &   3(-) &   1 &   0(+) &   0 \\
                  & 1.901 &  330 &  38(+) &  12 & 281(+) &  85 &  11(-) &   3 &   0(+) &   0 \\
\hline\hline
$\nres S31{1620}$ & 1.601 &  150 &  48(+) &  32 & 102(-) &  68 &     -- &  -- &     -- &  -- \\
                  & 1.601 &  152 &  51(+) &  34 & 101(-) &  66 &     -- &  -- &     -- &  -- \\
                  & 1.582 &  162 &  33(+) &  20 & 129(-) &  80 &     -- &  -- &     -- &  -- \\
\hline
$\nres P33{1232}$ & 1.229 &  113 & 113(+) & 100 &     -- &  -- &     -- &  -- &     -- &  -- \\
                  & 1.229 &  113 & 113(+) & 100 &     -- &  -- &     -- &  -- &     -- &  -- \\
                  & 1.230 &  113 & 113(+) & 100 &     -- &  -- &     -- &  -- &     -- &  -- \\[0.5ex]
$\nres P33{1600}$ & 1.675 &  406 &  52(+) &  13 & 354(+) &  87 &     -- &  -- &     -- &  -- \\
                  & 1.668 &  381 &  50(+) &  13 & 331(+) &  87 &     -- &  -- &     -- &  -- \\
                  & 1.674 &  384 &  50(+) &  13 & 334(+) &  87 &     -- &  -- &     -- &  -- \\
\hline
$\nres D33{1700}$ & 1.678 &  564 &  72(+) &  13 & 492(+) &  87 &     -- &  -- &     -- &  -- \\
                  & 1.678 &  512 &  68(+) &  13 & 444(+) &  87 &     -- &  -- &     -- &  -- \\
                  & 1.680 &  541 &  70(+) &  13 & 471(+) &  87 &     -- &  -- &     -- &  -- \\
\end{tabular} 
\end{center}
\caption{Extracted resonance parameters using KA84. First line:
  KA84-pp, second: KA84-ee, third: KA84-pt. $^a$: the coupling 
$g_{\eta N R}$ is given instead of the partial width, $^b$: width in 
keV. The signs of the couplings are given in brackets.}
\label{rescouplKA84}
\end{table}

\begin{table}[ht]
\begin{center}
\begin{tabular}{c|r|r|r|r|r|r|r|r|r|r}
                 & $M$ & $\Gamma_{tot}$ &
\multicolumn{2}{c|}{$\Gamma_{\pi N}$}  & \multicolumn{2}{c|}{$\Gamma_{\zeta N}$} & \multicolumn{2}{c|}{$\Gamma_{\eta N}$} & \multicolumn{2}{c}{$\Gamma_{K \Lambda}$} \\
$L_{2I,2S}$       &[GeV]  & [MeV]&  [MeV] & \% &  [MeV] & \% &  [MeV] & \% &  [MeV] & \% \\
\hline\hline
$\nres S11{1535}$ & 1.547 &  196 &  73(+) &  37 &  15(+) &   8 & 108(+) &  55 &   0(+) &   0 \\
                  & 1.544 &  156 &  63(+) &  40 &   9(+) &   6 &  84(+) &  54 &   0(+) &   0 \\
                  & 1.543 &  151 &  56(+) &  37 &   5(+) &   3 &  90(+) &  60 &   0(+) &   0 \\[0.5ex]
$\nres S11{1650}$ & 1.689 &  234 & 173(+) &  74 &  48(+) &  21 &   1(-) &   1 &  13(+) &   6 \\
                  & 1.687 &  213 & 157(+) &  74 &  45(+) &  21 &   0(-) &   0 &  11(+) &   5 \\
                  & 1.692 &  209 & 155(+) &  74 &  41(+) &  20 &   0(-) &   0 &  13(+) &   6 \\
\hline
$\nres P11{1440}$ & 1.463 &  400 & 252(+) &  63 & 148(+) &  37 & 2.37$^a$ &   0 &   0(+) &   0 \\
                  & 1.474 &  449 & 288(+) &  64 & 161(+) &  36 & 4.43$^a$ &   0 &   0(+) &   0 \\
                  & 1.448 &  334 & 202(+) &  60 & 132(+) &  40 & 0.95$^a$ &   0 &   0(+) &   0 \\[0.5ex]
$\nres P11{1710}$ & 1.714 &  195 &   0(+) &   0 &  97(-) &  50 &  69(+) &  35 &  29(+) &  15 \\
                  & 1.700 &  142 &   0(+) &   0 &  83(-) &  58 &  40(+) &  28 &  19(+) &  13 \\
                  & 1.727 &  266 &   1(+) &   0 & 138(-) &  52 &  89(+) &  33 &  38(+) &  14 \\
\hline
$\nres P13{1720}$ & 1.772 &  340 &  76(+) &  22 & 223(+) &  66 &  37(+) &  11 &   4(+) &   1 \\
                  & 1.766 &  348 &  77(+) &  22 & 241(+) &  69 &  25(+) &   7 &   5(+) &   1 \\
                  & 1.771 &  344 &  74(+) &  22 & 241(+) &  70 &  24(+) &   7 &   5(+) &   1 \\
\hline
$\nres D13{1520}$ & 1.508 &   92 &  51(+) &  55 &  41(-) &  45 & 16$^b$(+) &   0 &   0(+) &   0 \\
                  & 1.508 &   94 &  53(+) &  56 &  41(-) &  44 & 25$^b$(+) &   0 &   0(+) &   0 \\
                  & 1.510 &  101 &  58(+) &  57 &  43(-) &  43 & 10$^b$(+) &   0 &   0(+) &   0 \\[0.5ex]
$\nres D13{1700}$ & 1.909 &  352 &  40(+) &  11 & 289(+) &  82 &  23(-) &   7 &   0(+) &   0 \\
                  & 1.882 &  217 &  25(+) &  12 & 171(+) &  79 &  21(-) &  10 &   0(+) &   0 \\
                  & 1.901 &  359 &  35(+) &  10 & 300(+) &  83 &  24(-) &   7 &   0(+) &   0 \\
\hline\hline
$\nres S31{1620}$ & 1.595 &  148 &  42(+) &  28 & 106(-) &  72 &     -- &  -- &     -- &  -- \\
                  & 1.611 &  159 &  58(+) &  36 & 101(-) &  64 &     -- &  -- &     -- &  -- \\
                  & 1.598 &  150 &  44(+) &  29 & 106(-) &  71 &     -- &  -- &     -- &  -- \\
\hline
$\nres P33{1232}$ & 1.229 &  110 & 110(+) & 100 &     -- &  -- &     -- &  -- &     -- &  -- \\
                  & 1.230 &  110 & 110(+) & 100 &     -- &  -- &     -- &  -- &     -- &  -- \\
                  & 1.230 &  110 & 110(+) & 100 &     -- &  -- &     -- &  -- &     -- &  -- \\[0.5ex]
$\nres P33{1600}$ & 1.690 &  431 &  60(+) &  14 & 371(+) &  86 &     -- &  -- &     -- &  -- \\
                  & 1.685 &  440 &  62(+) &  14 & 378(+) &  86 &     -- &  -- &     -- &  -- \\
                  & 1.686 &  405 &  59(+) &  15 & 346(+) &  85 &     -- &  -- &     -- &  -- \\
\hline
$\nres D33{1700}$ & 1.689 &  661 &  85(+) &  13 & 576(+) &  87 &     -- &  -- &     -- &  -- \\
                  & 1.686 &  669 &  88(+) &  13 & 581(+) &  87 &     -- &  -- &     -- &  -- \\
                  & 1.675 &  547 &  70(+) &  13 & 477(+) &  87 &     -- &  -- &     -- &  -- \\
\end{tabular} 
\end{center}
\caption{Extracted resonance parameters using SM95. First line:
  SM95-pp, second: SM95-ee, third: SM95-pt. $^a$: the coupling 
$g_{\eta N R}$ is given instead of the partial width, $^b$: width in
keV. The signs of the couplings are given in brackets.}
\label{rescouplSM95}
\end{table}

\begin{table}[ht]
\begin{center}
\begin{tabular}{c|r|r|r|r|r|r|r|r}
                  & \multicolumn{4}{c|}{KA84} & \multicolumn{4}{c}{SM95} \\
\cline{2-9}
                  & $z_{\pi N}$ & $z_{\zeta N}$ & $z_{\eta N}$ & $z_{K \Lambda}$ &
                    $z_{\pi N}$ & $z_{\zeta N}$ & $z_{\eta N}$ & $z_{K \Lambda}$ \\
\hline\hline
$\nres P13{1720}$ &   1.440 &   0.216 &   0.348 &  -0.683 &
                     -1.771 &  -0.126 &  -1.375 &  -0.248 \\
                  &   1.150 &   0.180 &   0.877 &  -0.865 &
                     -0.379 &   0.142 &  -2.597 &  -1.471 \\
                  &  -1.013 &  -0.177 &  -1.207 &  -0.981 &
                     -2.200 &  -0.210 &  -1.993 &  -0.421 \\
\hline
$\nres D13{1520}$ &  -0.601 &   0.399 &  -1.383 &      -- &
                      0.423 &  -0.653 &   0.783 &      -- \\
                  &  -0.558 &   0.070 &  -1.005 &      -- &
                      0.366 &  -0.559 &   0.724 &      -- \\
                  &  -0.565 &   0.122 &  -1.135 &      -- &
                      0.352 &  -0.171 &   0.823 &      -- \\[0.5ex]
$\nres D13{1700}$ &   0.776 &   0.862 &   0.037 &  -0.749 &
                     -0.830 &   0.408 &  -0.079 &  -1.050 \\
                  &   0.523 &   0.722 &  -0.198 &  -0.536 &
                     -0.886 &  -1.113 &  -0.264 &  -1.980 \\
                  &  -0.396 &  -0.887 &  -0.689 &  -3.695 & 
                     -1.281 &  -0.990 &   0.195 &  -2.240 \\
\hline\hline
$\nres P33{1232}$ &  -0.333 &      -- &      -- &      -- &
                     -0.324 &      -- &      -- &      -- \\
                  &  -0.355 &      -- &      -- &      -- &
                     -0.354 &      -- &      -- &      -- \\
                  &  -0.383 &      -- &      -- &      -- &
                     -0.306 &      -- &      -- &      -- \\[0.5ex]
$\nres P33{1600}$ &   1.532 &   0.107 &      -- &      -- &
                      1.564 &   0.100 &      -- &      -- \\
                  &  -0.694 &  -0.006 &      -- &      -- &
                      0.844 &  -0.143 &      -- &      -- \\
                  &  -0.112 &  -0.765 &      -- &      -- &
                      1.587 &   0.094 &      -- &      -- \\
\hline
$\nres D33{1700}$ &   0.627 &  -0.215 &      -- &      -- &
                      0.588 &  -0.206 &      -- &      -- \\
                  &   0.628 &  -0.197 &      -- &      -- &
                     -0.725 &  -0.083 &      -- &      -- \\
                  &  -0.679 &   0.249 &      -- &      -- &
                      0.628 &  -0.212 &      -- &      -- \\
\end{tabular} 
\end{center}
\caption{Fitted $z$-parameters of the spin-$\dreih$-resonances.
  Notation as in Table \protect\ref{mesparmKA84SM95}.}
\label{reszparmKA84SM95}
\end{table}

\begin{table}[ht]
\begin{center}
\begin{tabular}{c|r|r|r|r}
                  &   $M$ & $\Gamma$ & $R \Gamma$ & $\Phi$ \\
                  & [GeV] &    [MeV] &      [MeV] & [$^\circ$] \\
\hline\hline
$\nres S11{1535}$ &        --$^a$ &        -- &        -- &   -- \\
                  &        --$^a$ &        -- &        -- &   -- \\
                  &         1.510 &       260 &       120 &   15 \\
                  &         1.487 &        -- &        -- &   -- \\
                  &         1.501 &       124 &        31 &  -12 \\[0.5ex]
$\nres S11{1650}$ & 1.660 - 1.669 & 137 - 166 &   30 - 40 &  -(38 - 48) \\
                  & 1.656 - 1.661 & 110 - 121 &   25 - 27 &  -(53 - 59) \\
                  &         1.640 &       150 &        60 &  -75 \\
                  &         1.670 &       163 &        39 &  -37 \\
                  & 1.673, 1.689$^b$ & 82, 192 &   22, 72 &  29, -85 \\
\hline
$\nres P11{1440}$ & 1.371 - 1.373 & 164 - 176 &   46 - 52 &  -(84 - 87) \\
                  & 1.357 - 1.362 & 143 - 155 &   37 - 42 &  -(94 - 95) \\
                  &         1.375 &       180 &        52 & -100 \\
                  &         1.385 &       164 &        40 &   -- \\
                  &         1.346 &       176 &        42 & -101 \\[0.5ex] 
$\nres P11{1710}$ & 1.674 - 1.690 &  82 - 150 &    5 - 11 &  80 - 94  \\
                  & 1.659 - 1.680 &  63 - 139 &    6 - 12 &  90 - 95 \\
                  &         1.690 &        80 &         8 &  175 \\
                  &         1.690 &       200 &        15 &   -- \\
                  &         1.770 &       378 &        37 & -167 \\
\hline
$\nres P13{1720}$ & 1.677 - 1.681 & 150 - 153 &   14 - 15 & -(115 - 120) \\
                  & 1.663 - 1.671 & 140 - 147 &   12 - 14 & -(116 - 120) \\
                  &         1.680 &       120 &         8 & -160 \\
                  &         1.686 &       187 &        15 &   -- \\
                  &         1.717 &       388 &        39 &  -70 \\
\hline
$\nres D13{1520}$ & 1.497 - 1.498 &   93 - 94 &        25 &  -(29 - 32) \\
                  &         1.496 &   86 - 94 &   24 - 28 &  -(28 - 30) \\
                  &         1.510 &       114 &        35 &  -12 \\
                  &         1.510 &       120 &        32 &   -8 \\
                  &         1.515 &       110 &        34 &    7 \\[0.5ex]
$\nres D13{1700}$ &        --$^a$ &        -- &        -- &   -- \\
                  &        --$^a$ &        -- &        -- &   -- \\
                  &         1.660 &        90 &         6 &    0 \\
                  &         1.700 &       120 &         5 &   -- \\
                  &            -- &        -- &        -- &   -- \\
\end{tabular} 
\end{center}
\caption{Values for the resonance poles and residues for the 
$I = \einh$-resonances compared to the results of other calculations. 
Shown are the range values of the three fits using KA84 (first line)
and SM95 (second) together with the values of Cutkosky et al. 
\protect\cite{cfhk79}, H\"ohler \protect\cite{h93} and Arndt et al. 
\protect\cite{sm95} in the following lines, respectively. $^a$: pole 
positions could not be deduced from the speed plots, $^b$: Arndt et
al. find two distinct cases.}
\label{resparmpoles12}
\end{table}

\begin{table}[ht]
\begin{center}
\begin{tabular}{c|r|r|r|r}
                  &   $M$ & $\Gamma$ & $R \Gamma$ & $\Phi$ \\
                  & [GeV] &    [MeV] &      [MeV] & [$^\circ$] \\
\hline\hline
$\nres S31{1620}$ & 1.598 - 1.603 & 101 - 108 &   15 - 16 & -(105 - 113) \\
                  & 1.588 - 1.595 &  91 - 123 &   11 - 16 & -(108 - 113) \\
                  &         1.600 &       120 &        15 & -110 \\
                  &         1.608 &       116 &        19 &  -95 \\
                  &         1.585 &       104 &        14 & -121 \\
\hline
$\nres P33{1232}$ &         1.208 &   93 - 94 &        47 &  -(49 - 50) \\
                  & 1.209 - 1.210 &   92 - 93 &        46 &  -48 \\
                  &         1.210 &       100 &        53 &  -47 \\
                  &         1.209 &       100 &        50 &  -48 \\
                  &         1.211 &       100 &        38 &  -22 \\[0.5ex]
$\nres P33{1600}$ &        --$^a$ &        -- &        -- &   -- \\
                  &        --$^a$ &        -- &        -- &   -- \\
                  &         1.550 &       200 &        17 & -150 \\
                  &         1.550 &        -- &        -- &   -- \\
                  &         1.675 &       386 &        52 &   14 \\
\hline
$\nres D33{1700}$ & 1.590 - 1.593 & 144 - 146 &        10 & -(46 - 49) \\
                  & 1.582 - 1.591 & 150 - 163 &   11 - 12 & -(47 - 53) \\
                  &         1.675 &       220 &        13 &  -20 \\
                  &         1.651 &       159 &        10 &   -- \\
                  &         1.655 &       242 &        16 &  -12 \\
\end{tabular} 
\end{center}
\caption{Same as in Table \protect\ref{resparmpoles12} but for the 
$I = \dreih$-resonances. $^a$: pole positions could not be deduced
from the speed plots.}
\label{resparmpoles32}
\end{table}

\clearpage


\begin{figure}[ht]
  \centerline{ \includegraphics[width=14cm]{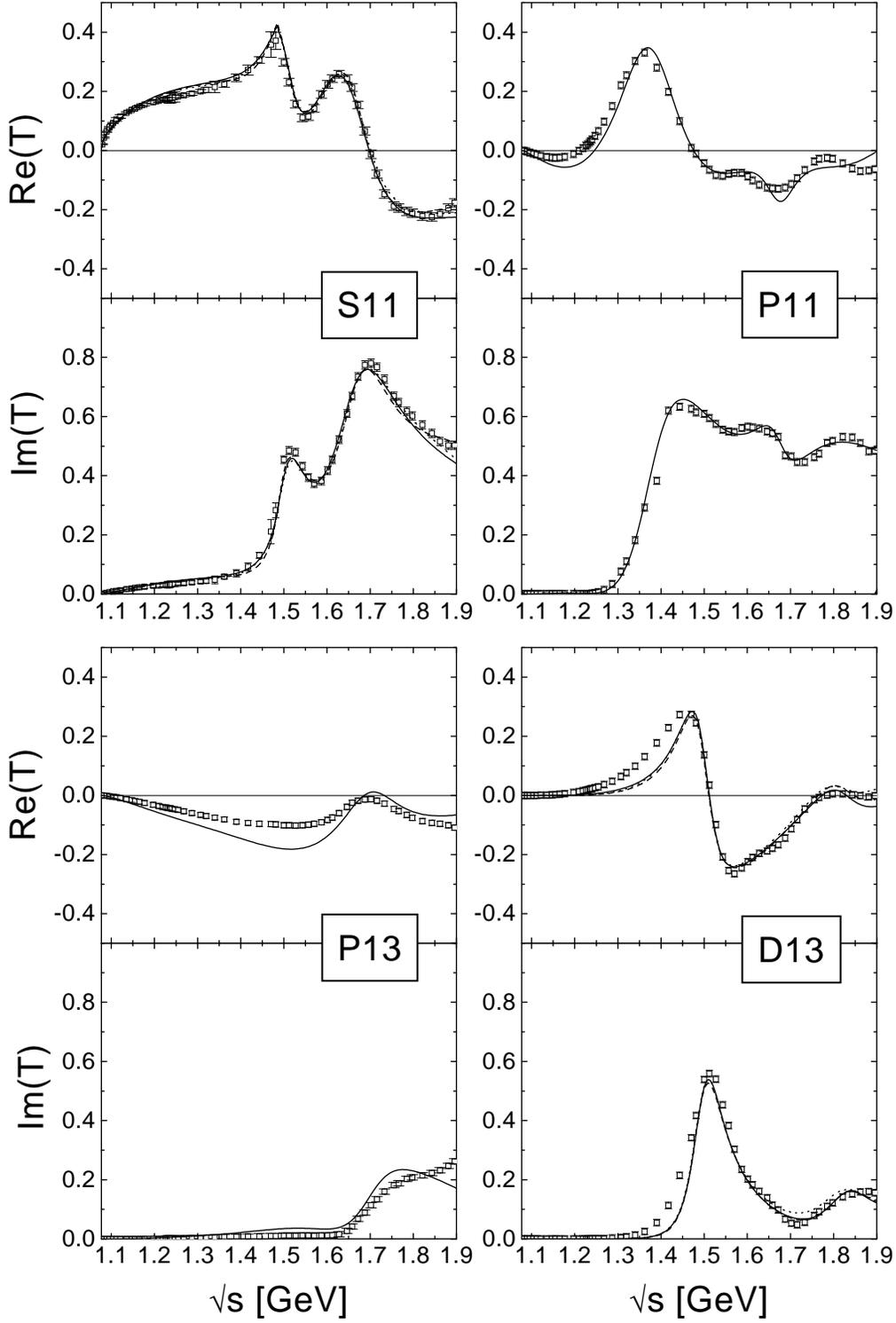} }
\caption{Fits to the $\pi N$ $I=\einh$-partial waves from KA84 
  \protect\cite{ka84}. Fit KA84-pt (\solid), KA84-pp (\dash) and
  KA84-ee (\dotdot). For Notation see Sec. \protect\ref{fitresults}.}
\label{ppi12KA84}
\end{figure}

\begin{figure}[ht]
  \centerline{ \includegraphics[width=14cm]{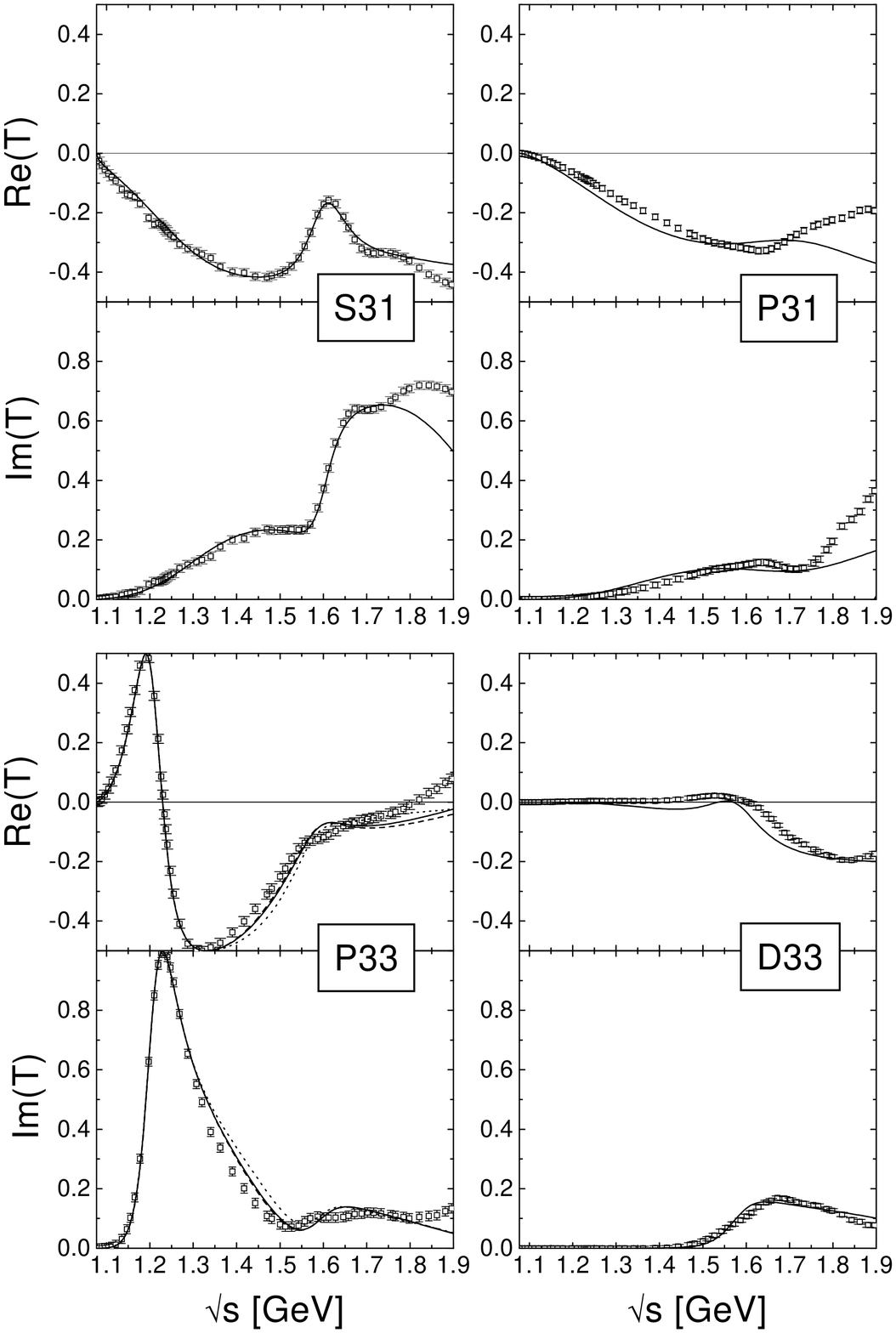} }
\caption{Results for the $I=\dreih$-channels. Legend as in Fig.\ 
  \protect\ref{ppi12KA84}.}
\label{ppi32KA84}
\end{figure}

\begin{figure}[ht]
  \centerline{ \includegraphics[width=14cm]{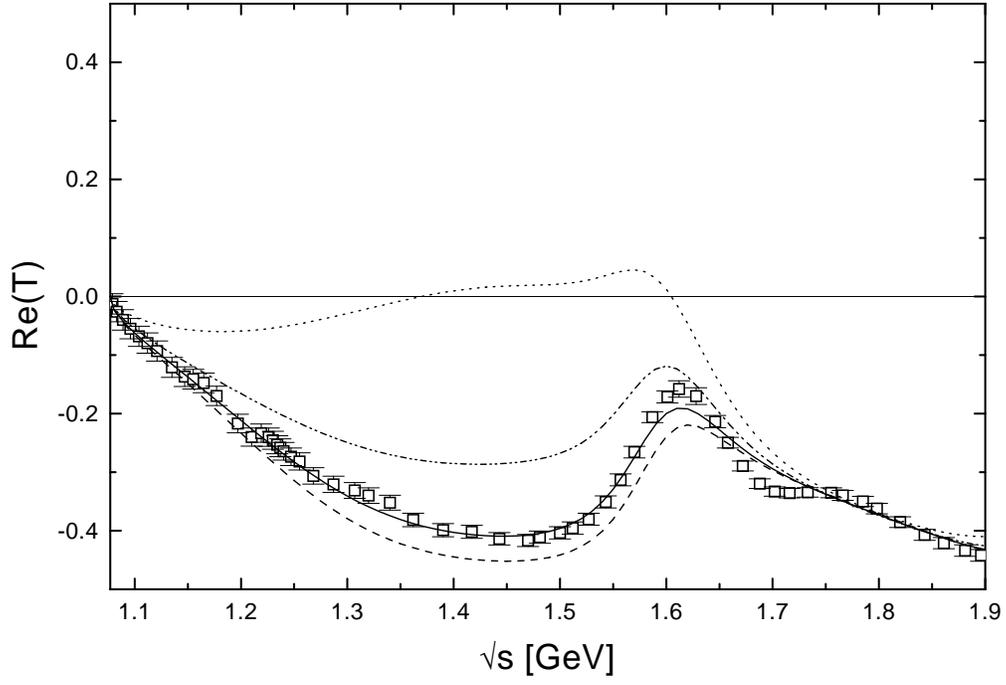} }
\caption{Influence of the $z_{\pi}$-parameter of the $\nres P33{1232}$
  on the $S_{31}$-phase shift. KA84-pt (\solid), $z_{\pi} = -0.5$
  (\dash), $z_{\pi} = 0.0$ (\dashdot), no $\nres P33{1232}$ (\dotdot).}
\label{pics31comp}
\end{figure}

\begin{figure}[ht]
  \centerline{ \includegraphics[width=14cm]{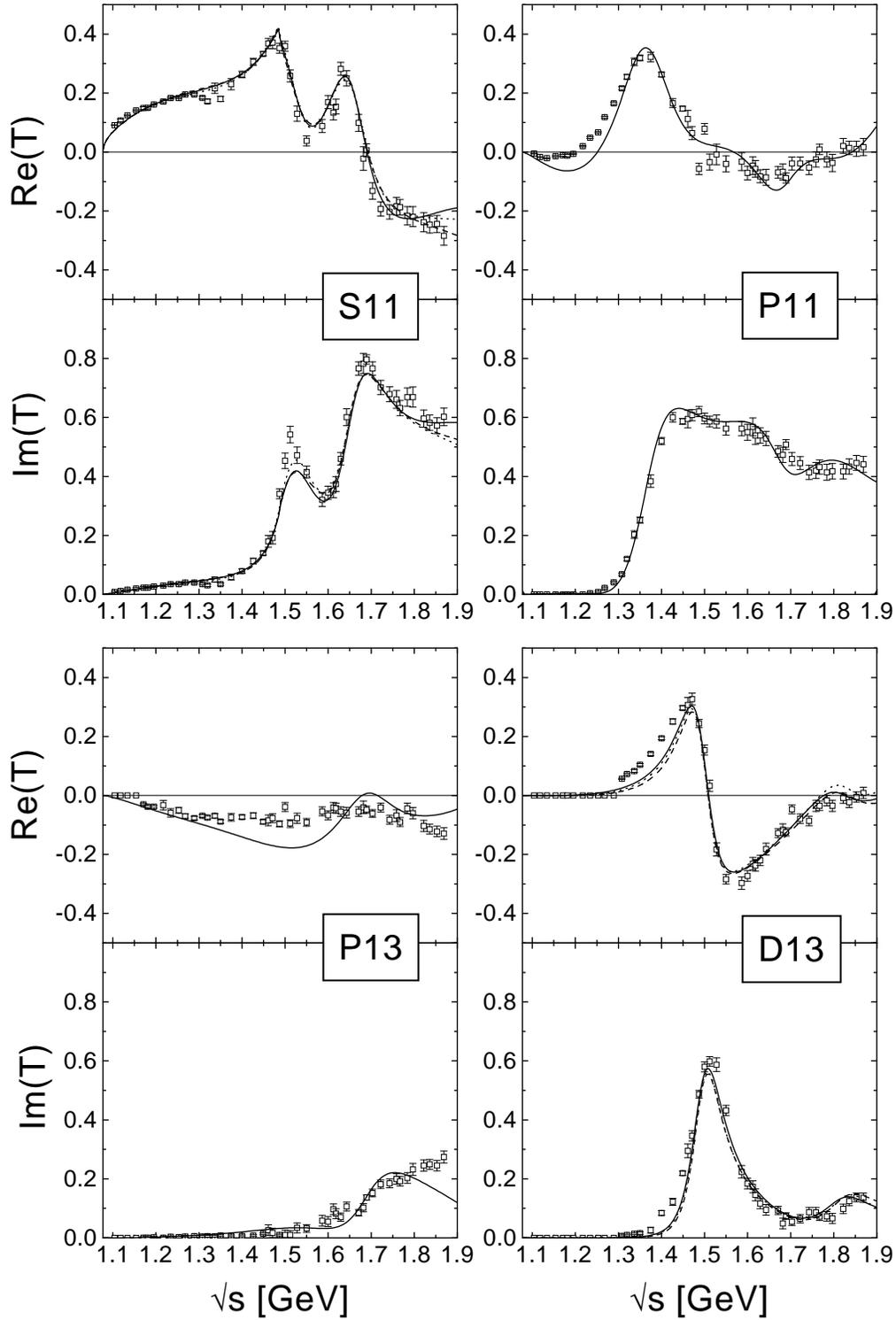} }
\caption{Fits to the $\pi N$ $I=\einh$-partial waves from SM95 
  \protect\cite{sm95}. Fit SM95-pt (\solid), SM95-pp (\dash) and
  SM95-ee (\dotdot).}
\label{ppi12SM95}
\end{figure}

\begin{figure}[ht]
  \centerline{ \includegraphics[width=14cm]{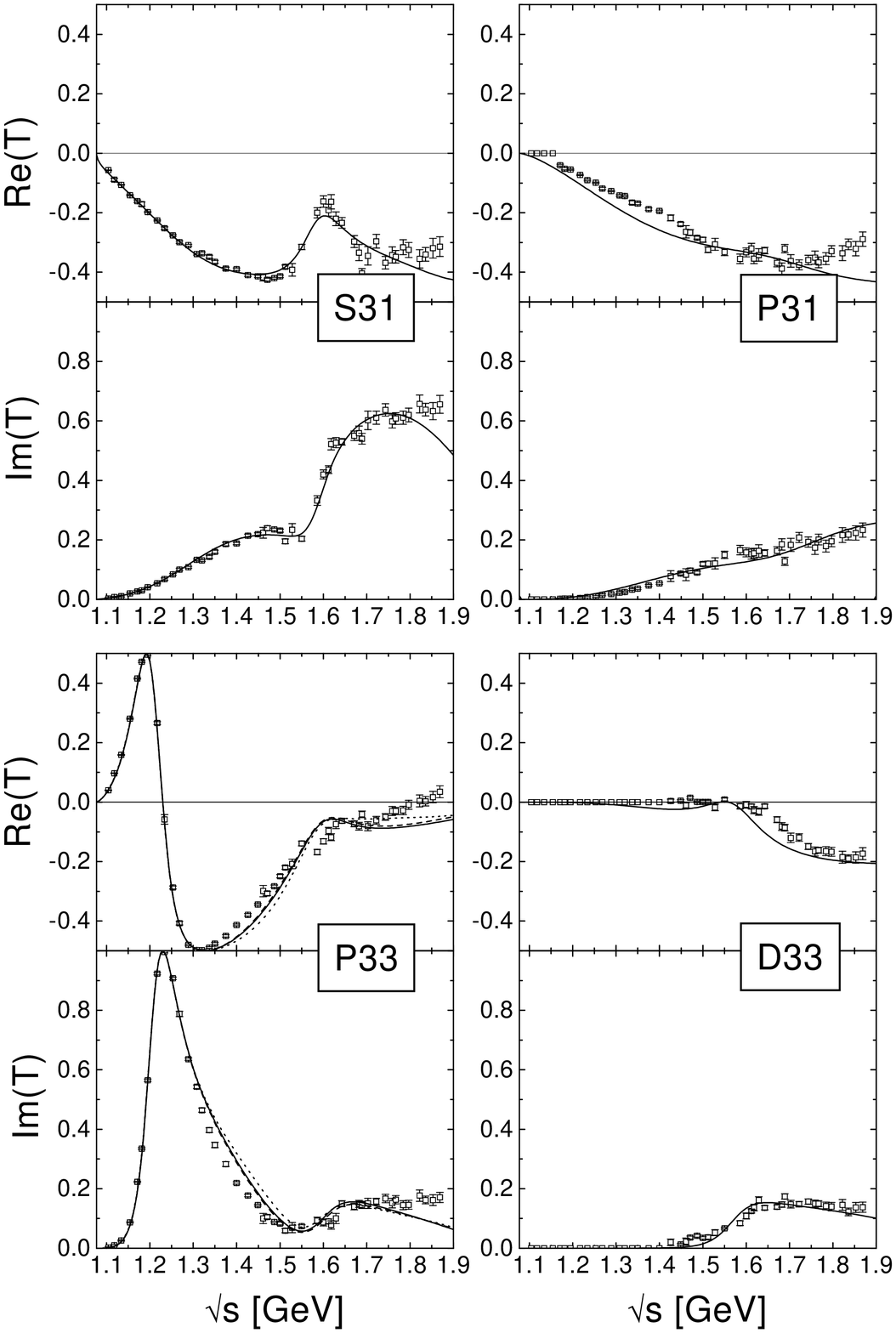} }
\caption{Results for the $I=\dreih$-channels. Legend as in Fig.\ 
  \protect\ref{ppi12SM95}.}
\label{ppi32SM95}
\end{figure}

\begin{figure}[ht]
  \centerline{ \includegraphics[width=14cm]{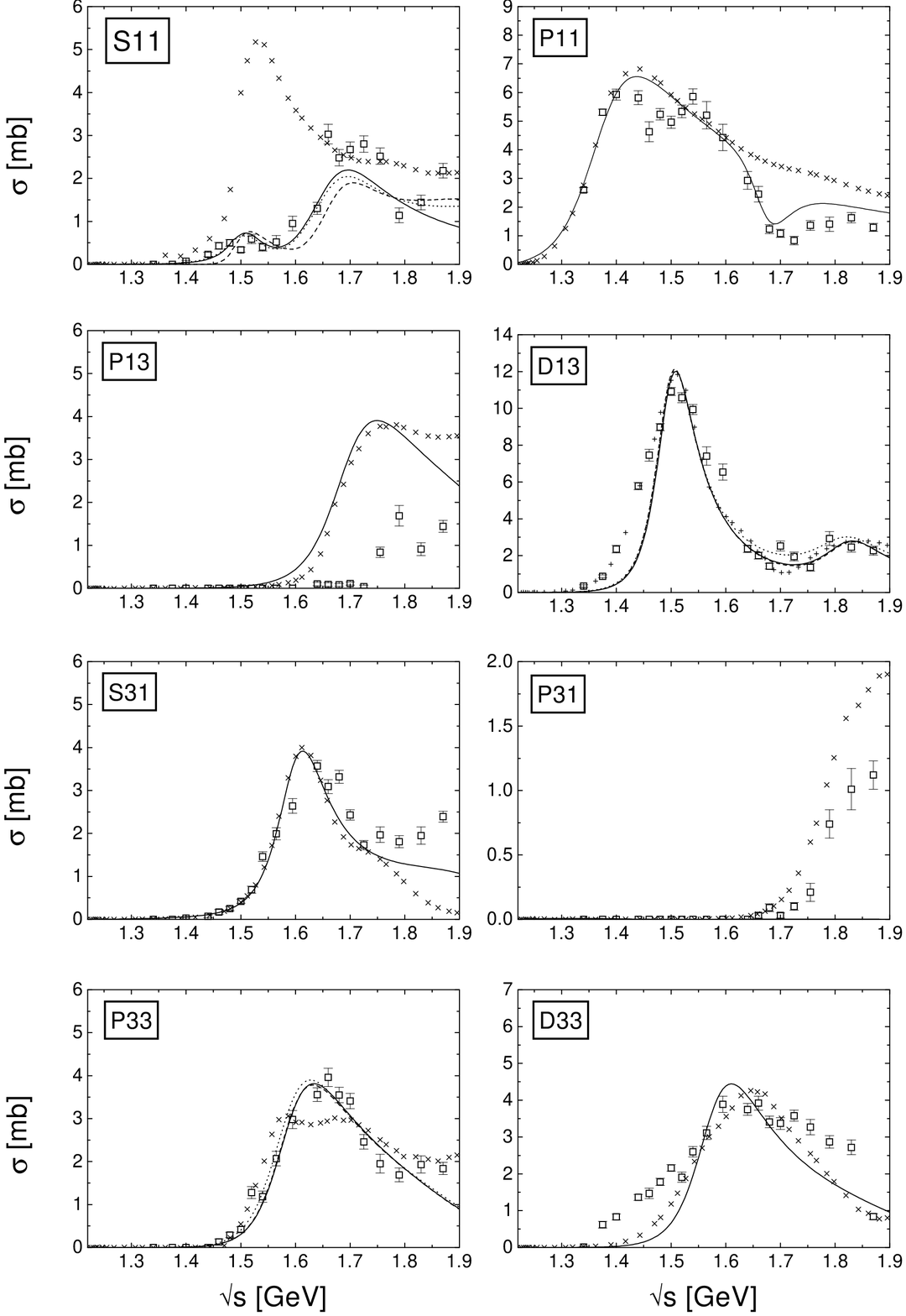} }
\caption{Comparison of the calculated $\pi N \to \pi\pi N$ cross
  sections for the fits using the KA84-PWA with the data from
  \protect\cite{ms92}. Legend as in Fig.\ \protect\ref{ppi12KA84}. In
  addition the inelastic cross section as determined from the KA84-PWA
  is shown ($\times$).}
\label{p2KA84}
\end{figure}

\begin{figure}[ht]
  \centerline{ \includegraphics[width=14cm]{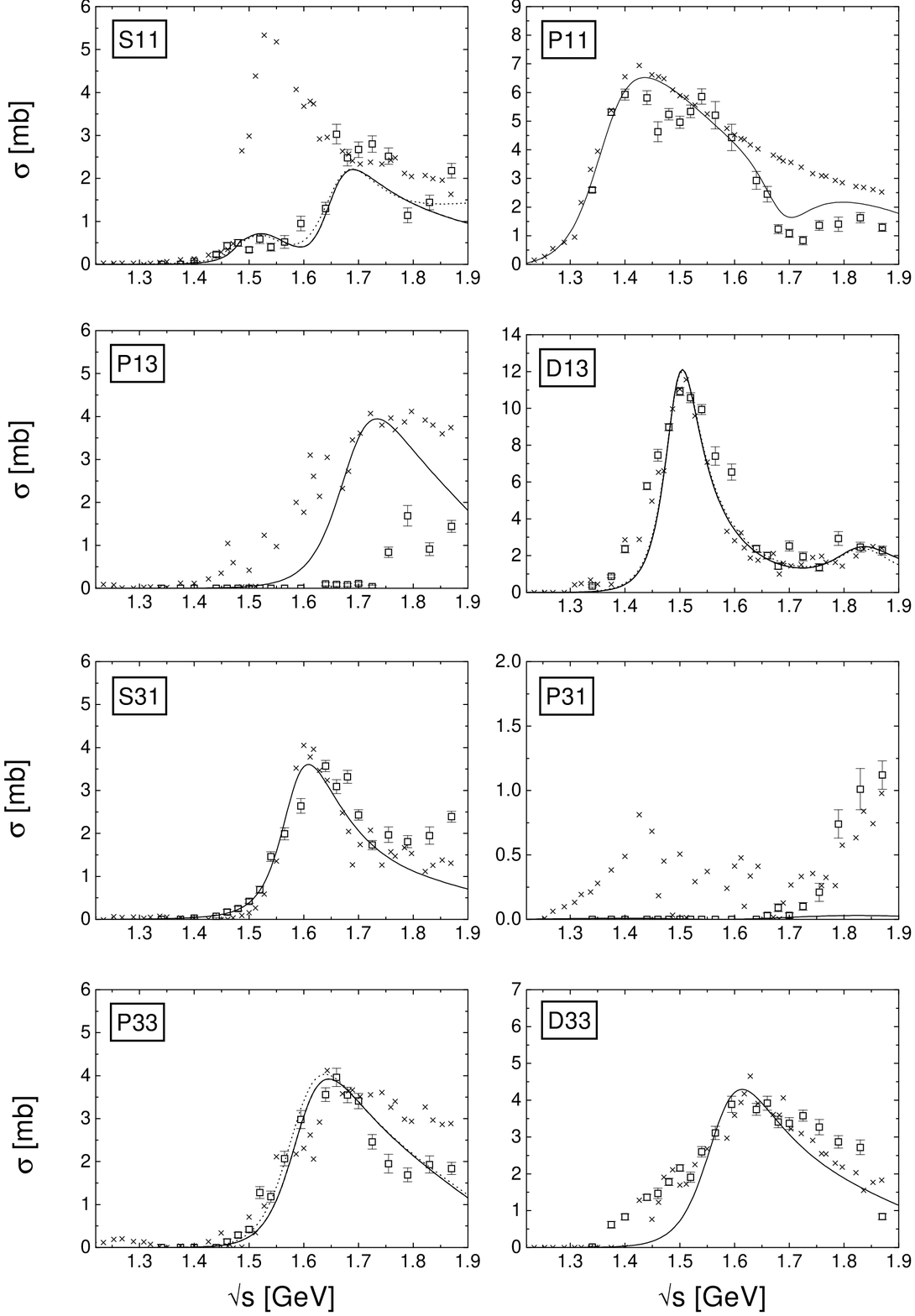} }
\caption{Comparison of the calculated $\pi N \to \pi\pi N$ cross
  sections for the fits using the SM95-PWA with the data from
  \protect\cite{ms92}. Legend as in Fig.\ \protect\ref{ppi12SM95}. In
  addition the inelastic cross section as determined from the SM95-PWA
  is shown ($\times$).}
\label{p2SM95}
\end{figure}

\begin{figure}[ht]
  \centerline{ \includegraphics[width=14cm]{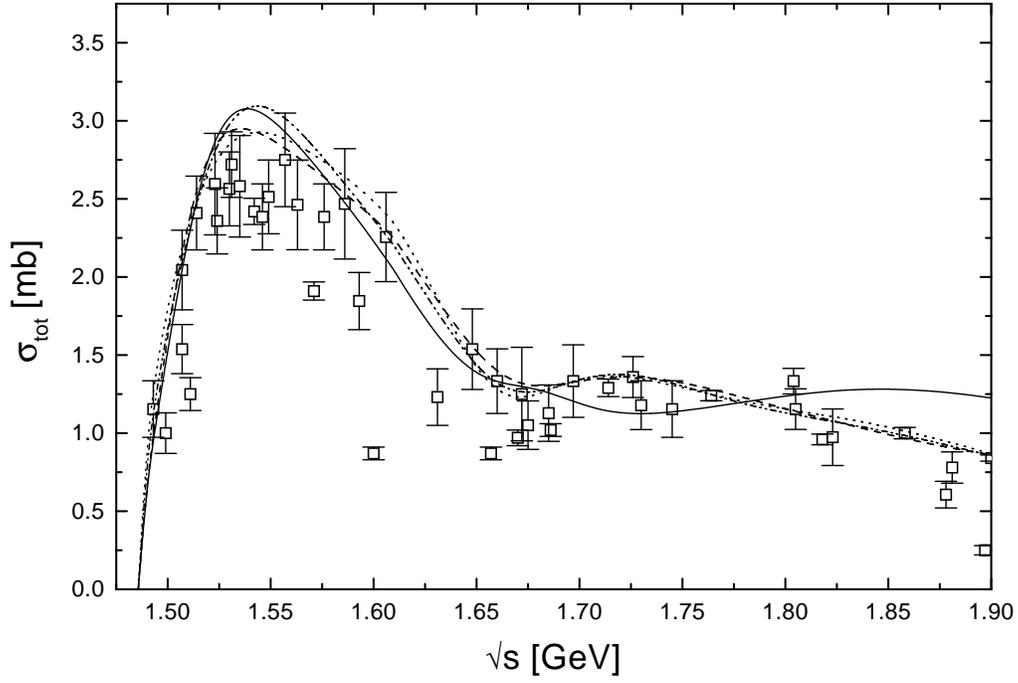} }
  \centerline{ \includegraphics[width=14cm]{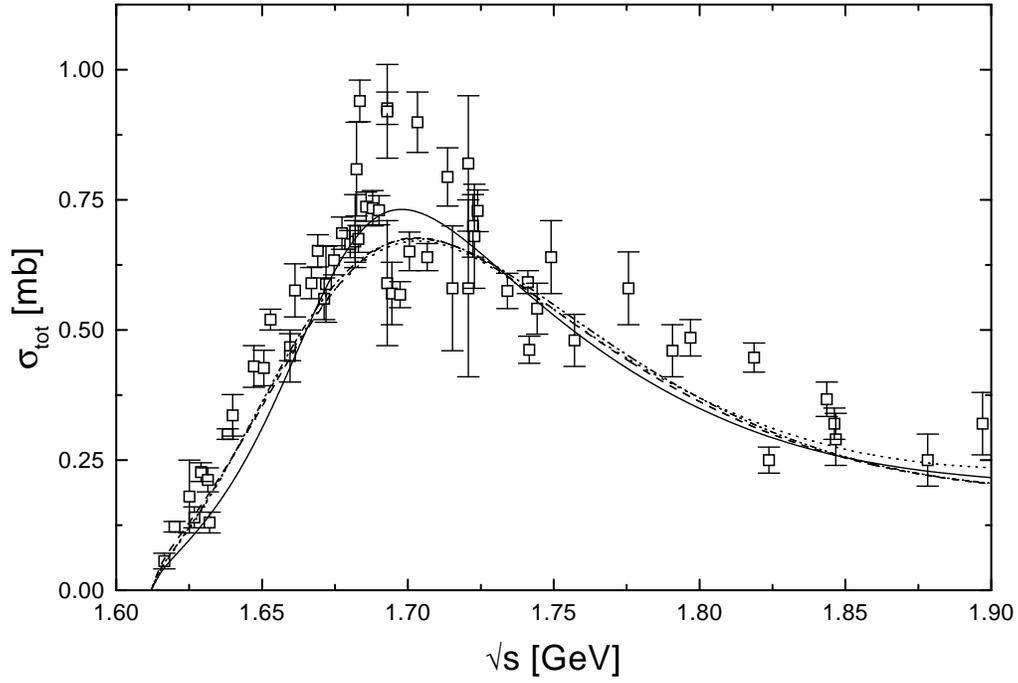} }
\caption{Results for the total $\pi^- p \to \eta n$ (upper plot) and 
  $\pi^- p \to K^0 \Lambda$ (lower) cross sections. Shown are the fits
  KA84-pt (\solid), KA84-pp (\dash), SM95-pt (\dashdot) and SM95-pp
  (\dotdot). Data as in Figs. \protect\ref{peKASM} and
  \protect\ref{pkKASM}.}
\label{pekKASM}
\end{figure}

\begin{figure}[ht]
  \centerline{ \includegraphics[width=14cm]{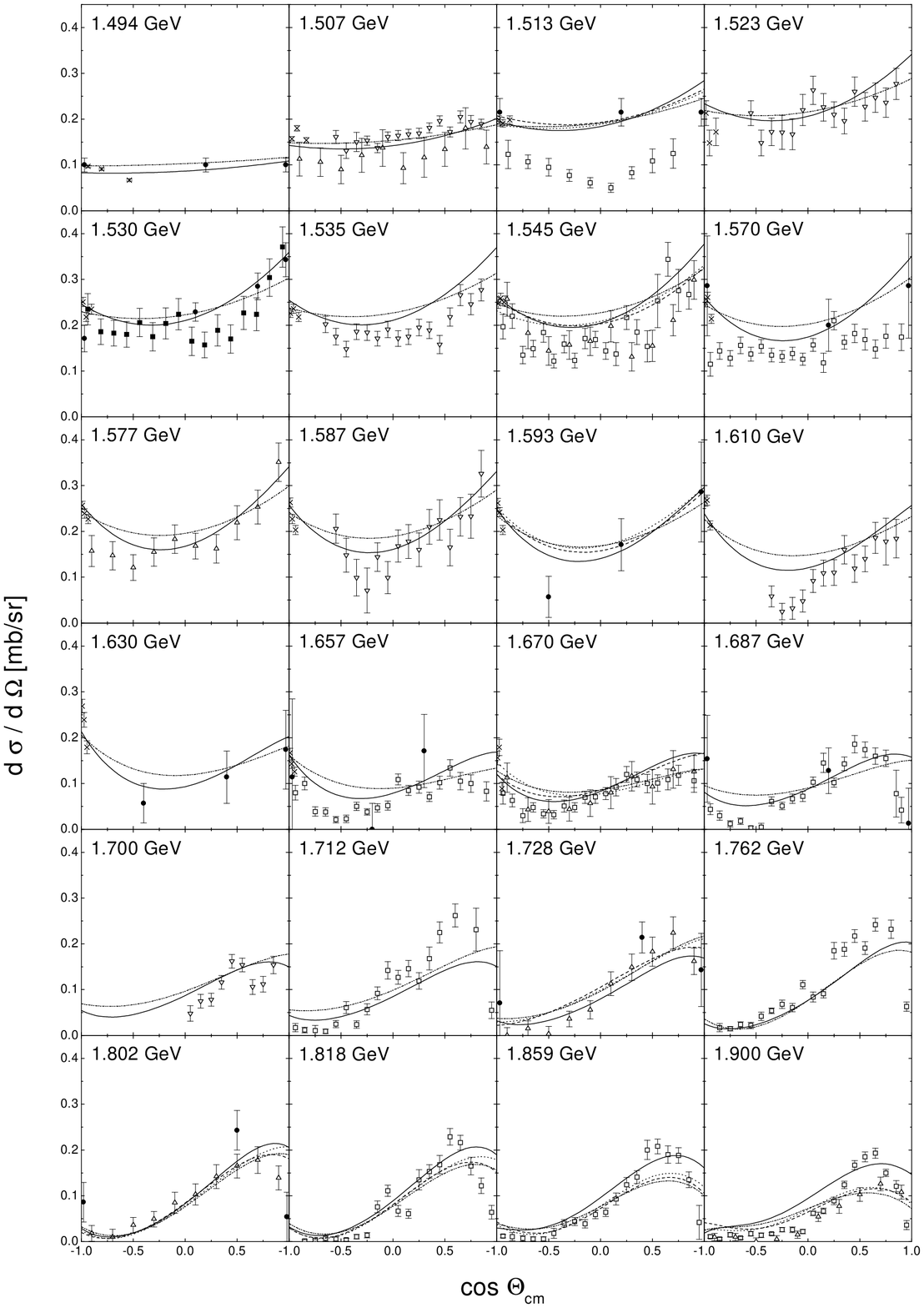} }
\caption{Comparison with data for the calculated differential 
  $\pi^- p \to \eta n$ cross sections. Legend as in Fig.\ 
  \protect\ref{pekKASM}. The datapoints are taken from:
  \protect\cite{bul69} ({\large $\bullet$}), \protect\cite{deb75}
  ($\times$), \protect\cite{dei69} ($\bigtriangledown$),
  \protect\cite{ric70} ($\bigtriangleup$), \protect\cite{bro79}
  ({\large $\Box$}), \protect\cite{fel75} (\FullBox).}
\label{peKASM}
\end{figure}

\begin{figure}[ht]
  \centerline{ \includegraphics[width=14cm]{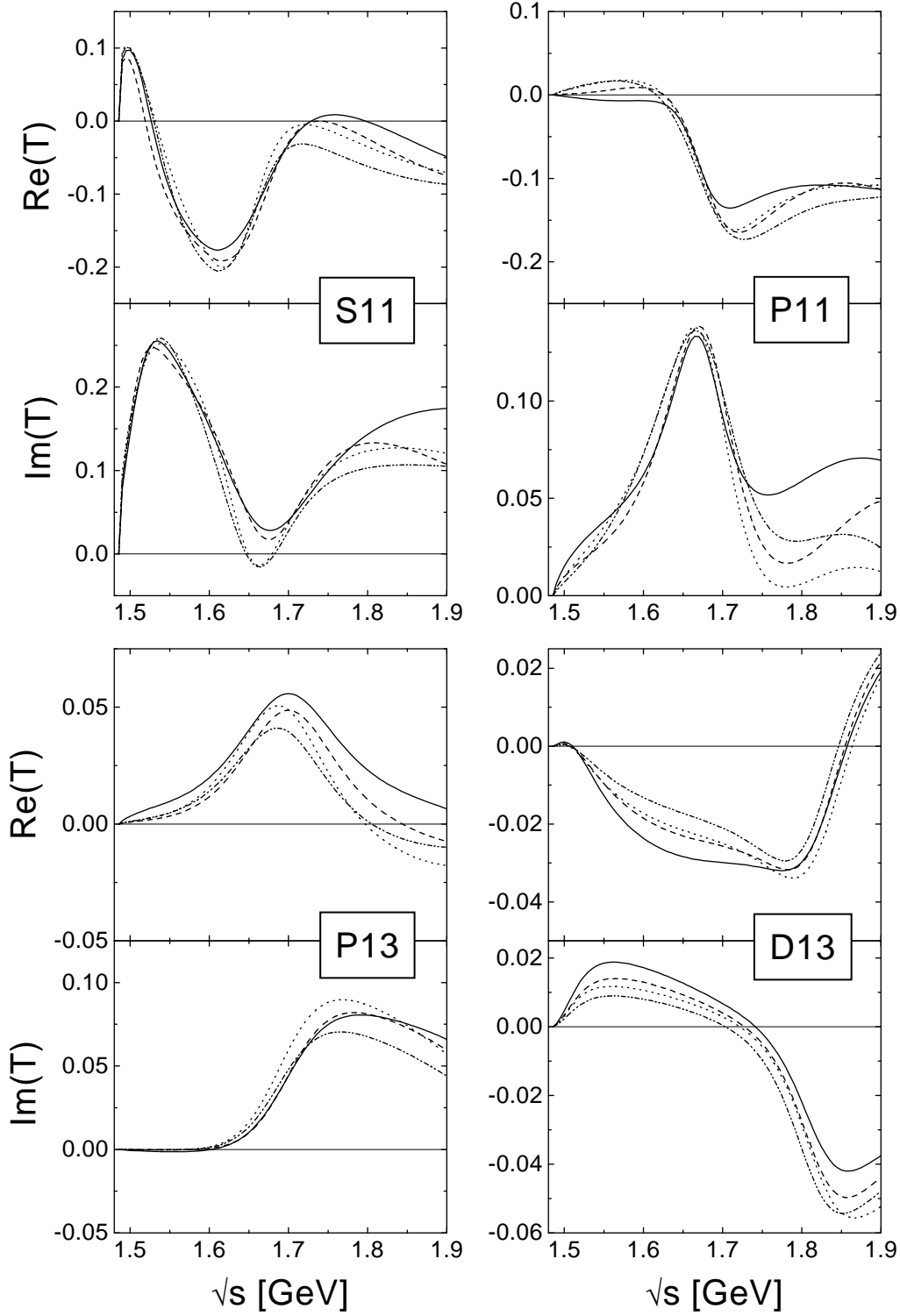} }
\caption{Calculated partial waves $T^{\einh}_{\pi \eta}$. Legend
  as in Fig.\ \protect\ref{pekKASM}.
}
\label{pei12KASM}
\end{figure}

\begin{figure}[ht]
  \centerline{ \includegraphics[width=14cm]{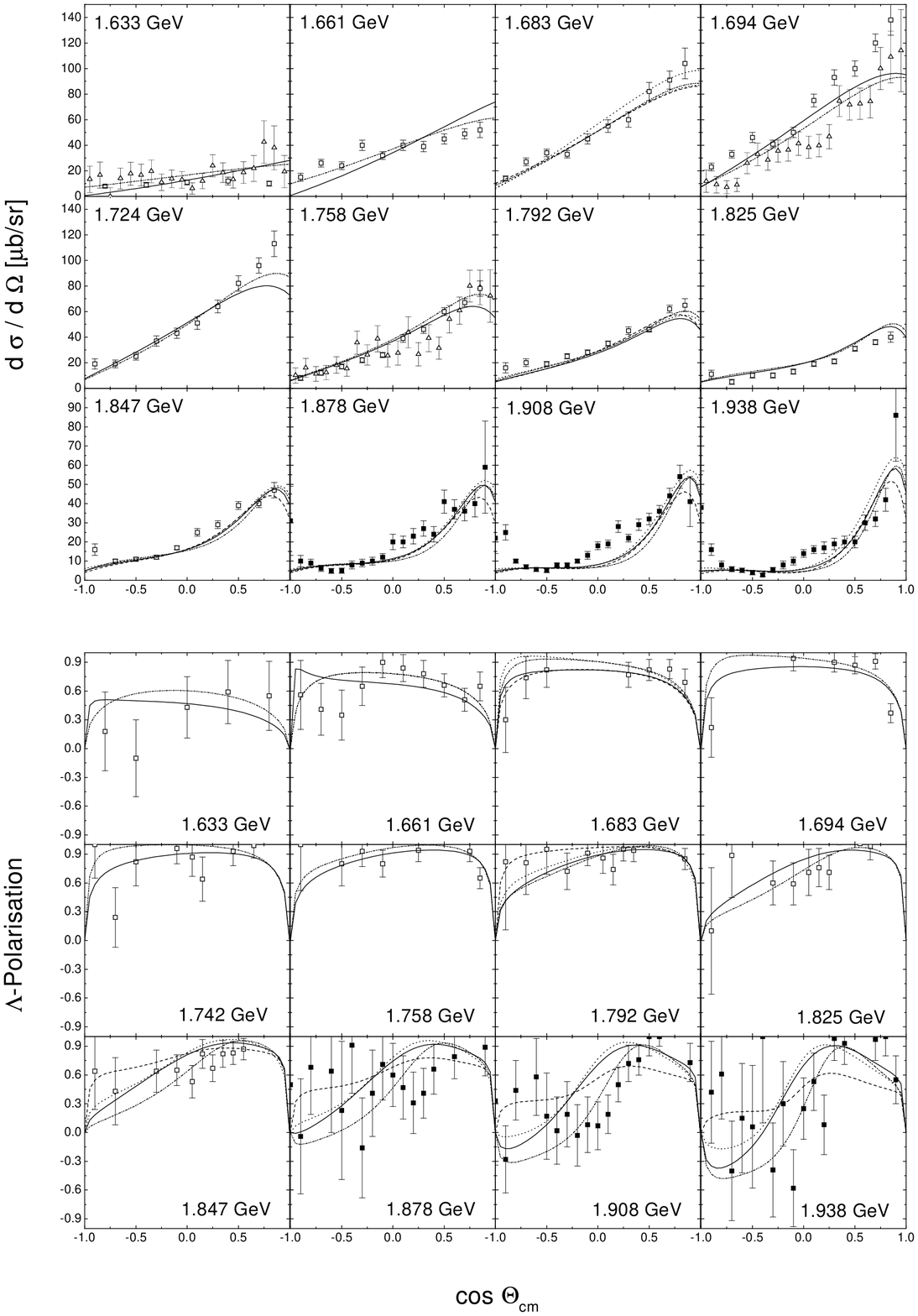} }
\caption{Comparison with data for the calculated differential 
$\pi^- p \to K^0 \Lambda$ cross sections and $\Lambda$-polarizations. 
Legend as in Fig.\ \protect\ref{pekKASM}. The datapoints are taken
from: \protect\cite{bak78} ({\large $\Box$}), \protect\cite{kna75} 
($\bigtriangleup$), \protect\cite{sax80} (\FullBox).}
\label{pkKASM}
\end{figure}

\begin{figure}[ht]
  \centerline{ \includegraphics[width=14cm]{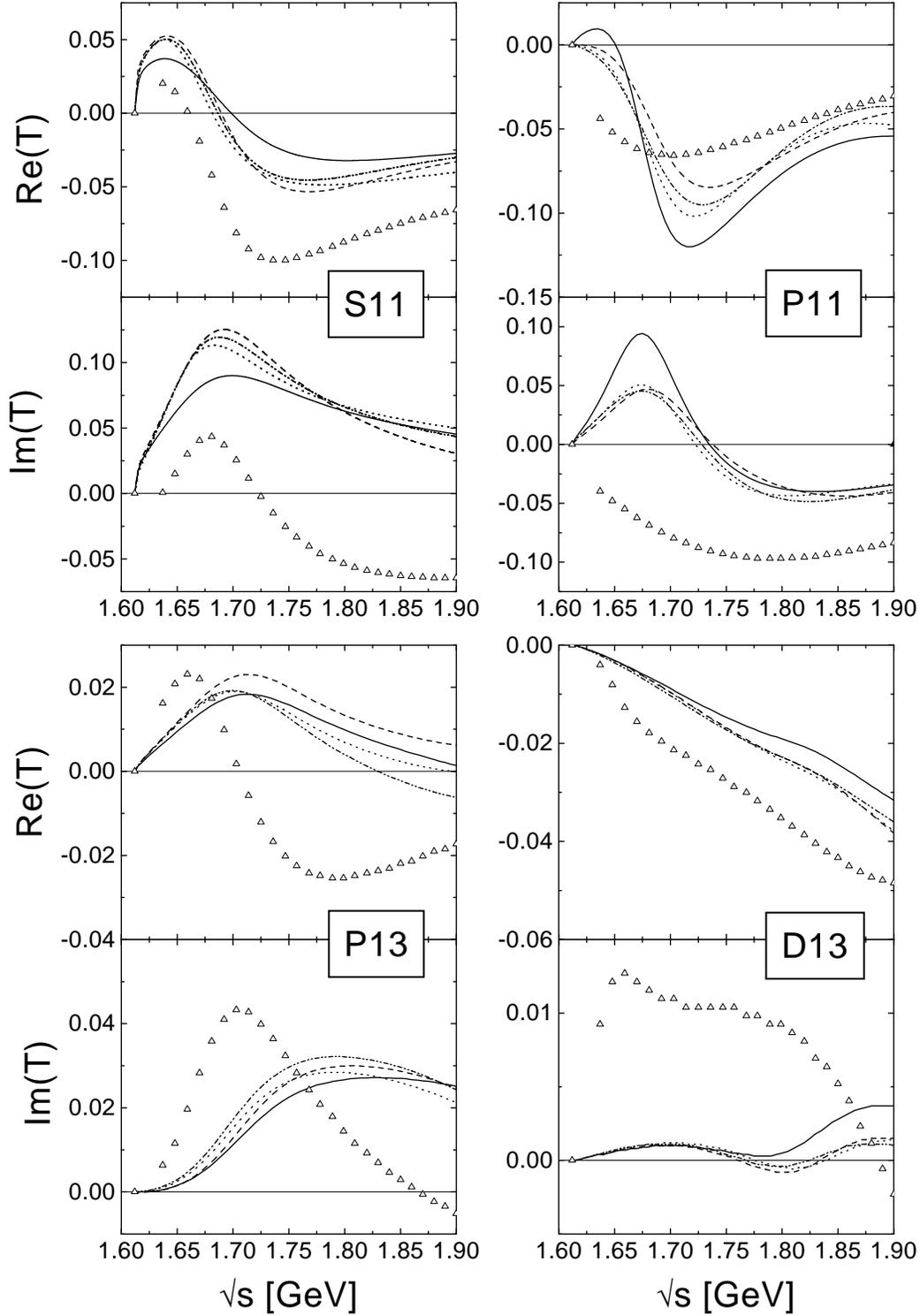} }
\caption{Calculated partial waves $T^{\einh}_{\pi K}$. Legend as
  in Fig.\ \protect\ref{pekKASM}. In comparison also the results of
  the calculation of Sotona and \v{Z}ofka \protect\cite{sz89}
  ($\bigtriangleup$) are shown.}
\label{pki12KASM}
\end{figure}

\begin{figure}[!ht]
  \centerline{ \includegraphics[width=14cm]{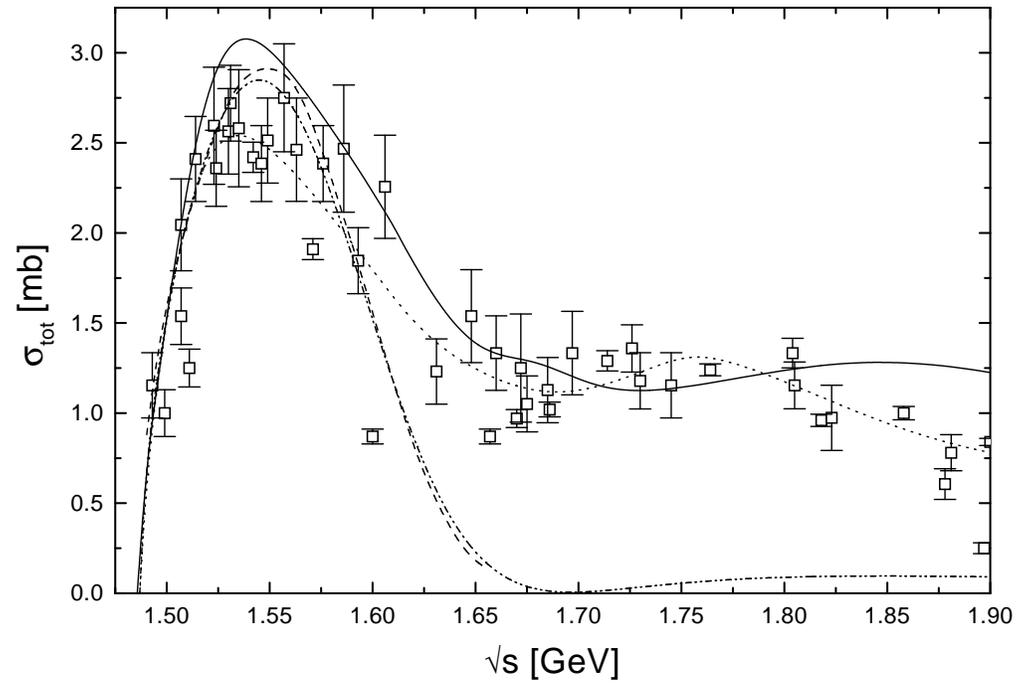} }
  \centerline{ \includegraphics[width=14cm]{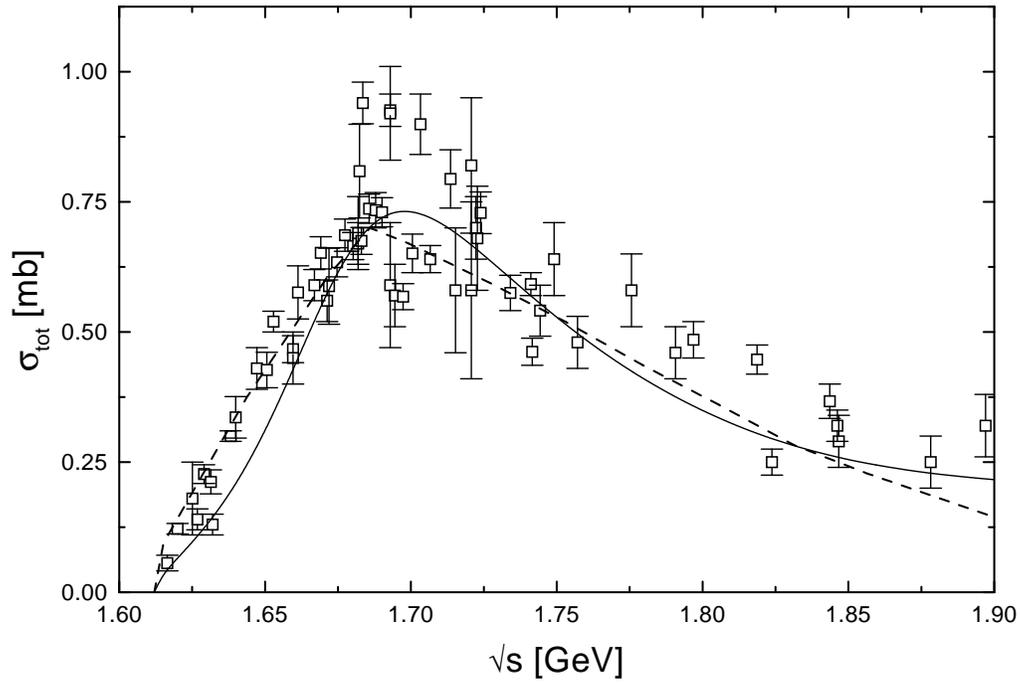} }
\caption{Comparison of the total $\pi^- p \to \eta n$ (upper plot) and
  $\pi^- p \to K^0 \Lambda$ (lower) cross sections. Fit KA84-pt
  (\solid), \protect\cite{kww97} (\dash), \protect\cite{sau96}
  (\dashdot), \protect\cite{bdssnl97} (\dotdot). Data as in Figs.
  \protect\ref{peKASM} and \protect\ref{pkKASM}.}
\label{pekcomp}
\end{figure}

\begin{figure}[ht]
  \centerline{ \includegraphics[width=14cm]{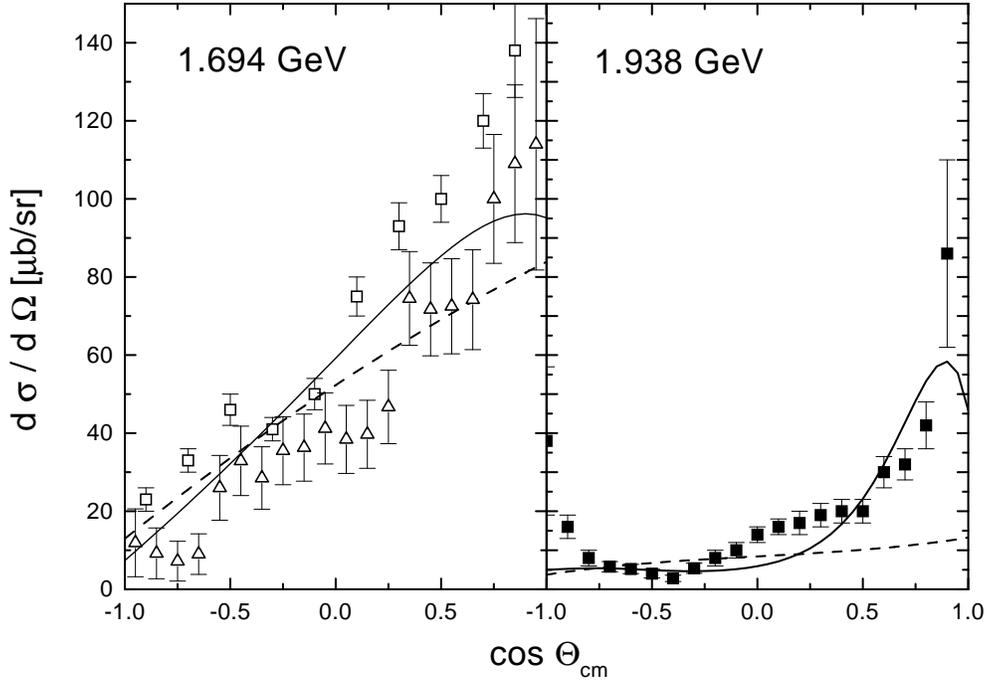} }
\caption{$K^*$-meson contribution to $\pi^- p \to K^0 \Lambda$ for two
  different energies. Shown is the fit KA84-pt with (\solid) and
  without (\dash) the $K^*$.}
\label{pickscomp}
\end{figure}

\begin{figure}[ht]
  \centerline{ \includegraphics[width=14cm]{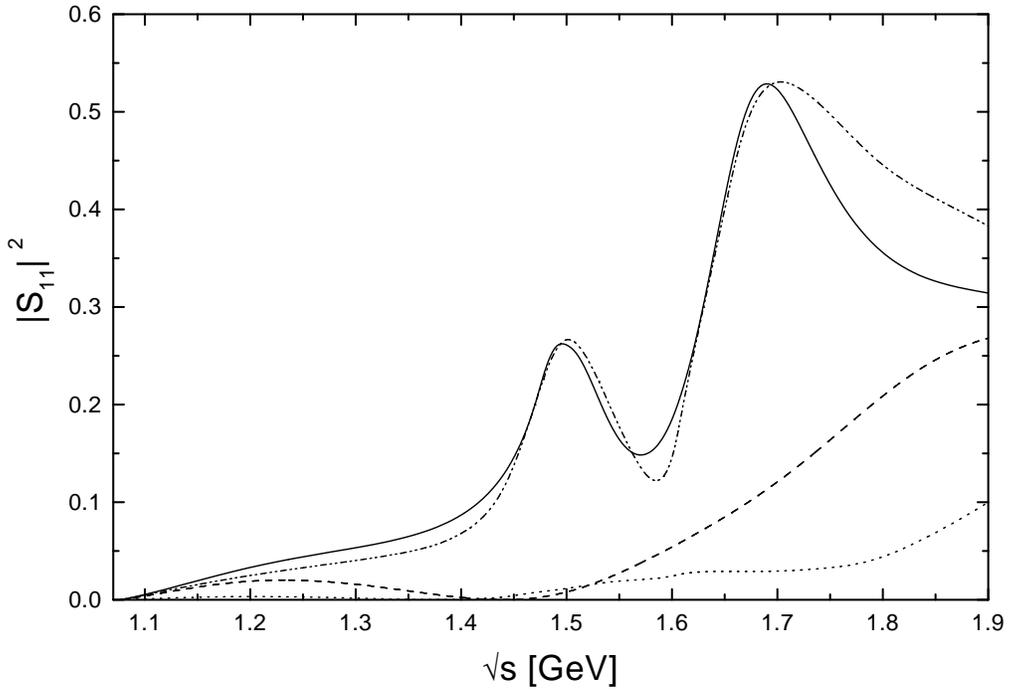} }
\caption{Comparison with the results from \protect\cite{dvl97}.
  Plotted is the square of the absolute value of the $S_{11}$-phase
  shift. Fit KA84-pt (\solid), background only (\dash), full
  calculation (\dashdot) and background (\dotdot) as given by 
\protect\cite{dvl97}.}
\label{pics11back}
\end{figure}

\begin{figure}[ht]
  \centerline{ \includegraphics[width=14cm]{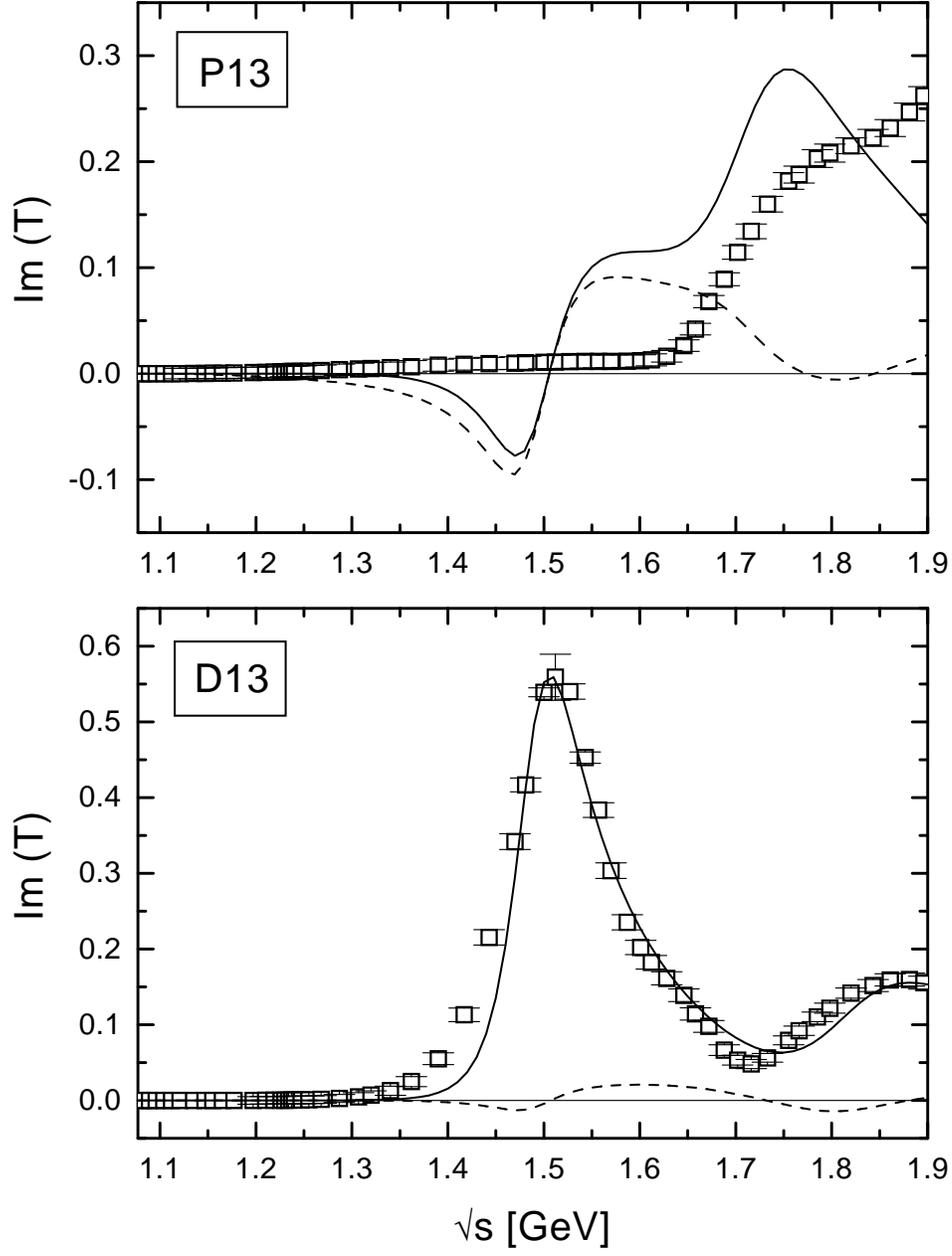} }
\caption{Results using the $T$-matrix approximation 
  (\protect\ref{tmatapprox}). Shown are the imaginary parts of the
  $P_{13}$- and $D_{13}$-partial waves for $\pi N$-scattering (\solid)
  and the corresponding values $\Delta T^{\alpha}$ (\dash). The data
  are from KA84 \protect\cite{ka84}.}
\label{unitarpp}
\end{figure}

\begin{figure}[ht]
  \centerline{ \includegraphics[width=14cm]{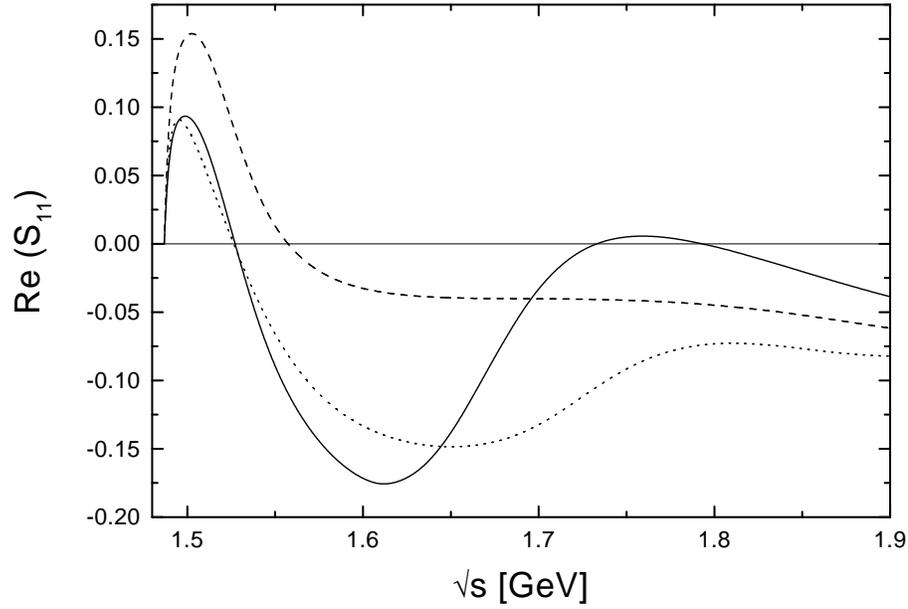} }
\caption{Influence of the $\nres S11{1650}$ on the $\pi N \to \eta
  N$ amplitudes. Shown is the real part of $S_{11}$ for the $K$-matrix
  calculation using KA84-pt with (\solid) and without (\dash) the
  $\nres S11{1650}$. For comparison we also show the $T$-matrix result
  (\dotdot).}
\label{unitarpe} \end{figure}
\end{document}